\documentclass[twocolumn]{aastex701}

\usepackage{amsmath}
\usepackage{amssymb}
\usepackage{euscript}
\usepackage{bbold}
\usepackage{paralist}
\usepackage{color}
\usepackage{amsmath}
\usepackage{lineno}
\usepackage{booktabs}
\usepackage{threeparttable}
\usepackage{adjustbox}
\usepackage{tabularx}
\usepackage{svg}
\usepackage{color}
\usepackage{enumitem}
\usepackage{graphicx}
\usepackage{comment}
\usepackage{placeins}
\usepackage{hyperref}
\begin{document}





\title{Mapping the Landscape of M Dwarf X-ray Flares: New Discoveries in Context}

\author[0009-0006-2117-8993]{Imri A. Dickstein}
\email{imri94@gmail.com}
\affiliation{Department of Physics, Bar-Ilan University, Israel}
\author[0000-0001-6753-1488]{Maayane T. Soumagnac}
\email{<imri.email@domain>}
\affiliation{Department of Physics, Bar-Ilan University, Israel}
\affiliation{Lawrence Berkeley National Laboratory, 1 Cyclotron Road, Berkeley, CA 94720, USA}
\author[0000-0002-6786-8774]{Eran O. Ofek}
\email{<imri.email@domain>}
\affiliation{Benoziyo Center for Astrophysics, Weizmann Institute of Science, 76100 Rehovot, Israel}
\author[0009-0001-3501-7852]{Param Rekhi}
\email{<imri.email@domain>}
\affiliation{Benoziyo Center for Astrophysics, Weizmann Institute of Science, 76100 Rehovot, Israel}
\author[0000-0002-6859-0882]{Volker Perdelwitz}
\email{<imri.email@domain>}
\affiliation{Department of Earth and Planetary Sciences, Weizmann Institute of Science, Rehovot 7610001, Israel}
\author[0000-0001-6760-3074]{Sagi Ben-Ami}
\email{<imri.email@domain>}
\affiliation{Benoziyo Center for Astrophysics, Weizmann Institute of Science, 76100 Rehovot, Israel}
\author[0000-0002-6540-1484]{Thomas Kupfer}
\email{<imri.email@domain>}
\affiliation{Hamburger Sternwarte, University of Hamburg, Gojenbergsweg 112, 21029 Hamburg, Germany}
\affiliation{Department of Physics \& Astronomy, Texas Tech University, Box 41051, Lubbock, TX, 79409-1051, USA}


\begin{abstract}

We report the discovery of $11$ X-ray flares from $7$ M dwarfs previously unknown to exhibit flaring activity, by cross-matching eROSITA observations of bright, nearby M dwarfs with the Chandra telescope archive. To analyze the properties of these flares in a broader context, we compile the sample of all reported X-ray flares from the $15$ M dwarfs identified as flaring in the literature. We use this combined sample to derive constraints on the X-ray flare frequency distributions of M0–M6 stars. The average flare occurrence rate we measure is $\sim 10^{-1}\,\rm ks^{-1}$ (corresponding to $\sim 9$ flares per day). The X-ray flares in this sample span energies from $10^{29}\,\rm erg$ to $10^{33}\,\rm erg$ and exhibit a strong correlation between flare strength and duration. The flare properties we characterize include their durations, flux and temperature enhancements, and temporal asymmetries.
Using these results and recent simulations of flare-driven atmospheric escape, we derive an upper limit on the time required for habitable Earth-like planets orbiting these M dwarfs to completely lose their atmospheres: $0.5$--$30\,\rm Myr$.

\end{abstract}


\section{Introduction}


The flaring properties of M dwarfs have been extensively studied in optical wavebands (e.g. \citealt{Kowalski2010, hunt2012,hawley2014, davenport2016, Yang2017,howard2019,Mondrik2019, Schmidt2019, Doyle2019, Martinez2020, Gunther2020, raetz2020, Murray2022}) and in the UV (e.g. \citealt{Welsh2007,hawley2007, Loyd2018, Loyd2018b, Froning2019,kowalski2019,chavali2022,fleming2022, Rekhi2023}). 

In the X-ray, $\sim120$ X-ray flares have been reported to date, with $\sim40\%$ of them originating from intensive observations of a single star, AU\,Mic \citep{tristan2023, pye2015,smith2005}. Apart from this star, only $14$ additional flaring M dwarfs have been documented, either through targeted X-ray observations of individual active stars (e.g., \citealt{Robrade2005}) 
or during coordinated multiwavelength campaigns targeting small samples of known flaring stars (e.g., \citealt{favata2000,mitrakraev2005,Mitra2005, osten2005, smith2005, fuhrmeister2007,osten2010, Liefke2010, fuhrmeister2011, namekata2020, Kuznetsov2021, Stelzer2022a, kuznetsov2023, tristan2023, inoue2024, ilin2024}). Notably, \citet{pye2015} adopted a more systematic approach by cross-matching the XMM-Newton Serendipitous Source Catalogue \citep{watson2009} with the Hipparcos–Tycho catalog \citep{hog2000}, which led to the identification of $23\%$ of the flares known to date and more than half of the flaring M dwarfs known to date (see \S~\ref{sec:lit}).

Increasing the sample of flaring M dwarfs and improving our understanding of their X-ray flaring behavior are important for answering several astrophysical questions. We list a few examples below.

Stellar flare at all wavelengths provide insight into uncharted physical processes taking place in stellar interiors. In particular, they are theorized to originate from (and probe the efficiency of) the stellar dynamo, the mechanism that generates the stellar magnetic field through the combined action of a near-surface convection zone and differential rotation \citep{katsova2018}.
The stellar dynamo process is believed to operate most effectively in stars on the cooler half of the Hertzsprung–Russell diagram \citep{stelzer2017}, especially M dwarfs, which exhibit larger convective regions than higher-mass stars \citep{reiners2012, hawley2014, newton2017} and span the transition from partially to fully convective interiors \citep{parker1993, chabrier1997, baraffe2018}. 
Additionally, flares can shed light on the role of magnetic reconnection in shaping coronal temperatures and densities \citep{gudel2004, cassak2008}. 

Stellar flares can impact the habitability of potential exoplanets. As the most numerous stars in our galaxy \citep{Covey2008, Bochanski2010}, M dwarfs are prime candidates for hosting habitable exoplanets \citep{Tarter2007, scalo2007, Shields2016}. They offer several observational advantages for exoplanet searches, including (1) the proximity of the habitable zone, which yields frequent planetary transits and (2) their small star-to-planet mass and size ratios, which produce strong signals in both transit and radial-velocity measurements. Moreover, planets surrounding M dwarfs are particularly favorable targets for atmospheric characterization, especially terrestrial-sized planets \citep{Kempton2018}. For all these reasons, M dwarfs have been primary targets in recent exoplanet search programs such as TESS, CARMENES, and MEarth \citep{Ricker2015, Ribas2023, Irwin2009}. However, violent flaring episodes in the UV and X-ray may jeopardize the habitability of planets orbiting M dwarfs. Several studies have shown that flaring at these wavelengths can influence planetary evolution \citep{Tarter2007, scalo2007, Stelzer2013, Shibata2016}, alter atmospheric chemical composition \citep{Stelzer2013}, cause atmospheric erosion \citep{Lammer2007}, trigger runaway greenhouse effects and lead to thermal and hydrodynamic atmospheric escape \citep{murrayclay2009, koskinen2013, lugera2015}. High UV/X-ray fluxes from stellar flares can also lead to the buildup of detectable levels of false positive biosignatures, further complicating assessments of planetary habitability \citep{grenfell2012, tian2015, venot2017, meadows2018}.

Finally, stellar flares act as contaminants in many large-scale transient surveys across a broad range of wavelengths. Transients surveys such as the Zwicky Transient Facility \citep[ZTF; e.g.,][]{voloshina2024} and the All-Sky Automated Survey for Supernovae \citep[ASAS-SN;][]{Schmidt2019}
, routinely encounter contamination from flares and have published extensive catalogs dedicated to identifying and characterizing M dwarf flaring behavior. A similar challenge will have to be overcome in future planned transients surveys conducted with ULTRASAT \citep{shvartzvald2024}, LS4 \citep{miller2025}, and LSST \citep{ivezi2019}, which makes the need to understand and characterize these objects all the more urgent. For example, ULTRASAT alone is expected to capture more than $4 \times 10^{5}$ M dwarf flares \citep{shvartzvald2024} over the course of the survey.

The study presented in this paper has two goals: (1) To discover additional flaring M dwarfs by leveraging the overlap between twenty-five years of Chandra archival data and the eROSITA surveys \citep{weisskopf2000, predehl2021}. (2) To obtain an initial picture of the parameter space of X-ray flares from M dwarfs, by assembling the properties of previously reported flares and the new events presented here into a single framework.


This paper is organized as follows. In \S\ref{sec:dataset}, we describe the sample of X-ray flaring stars we compiled. In \S\ref{sec:flares_identifications}, we present our methodology for identifying X-ray flares in Chandra observations. In \S\ref{sec:flares_characterization}, we describe the properties of the X-ray flares and in \S\ref{results} we analyze the properties of the stars hosting these flares and derive a constraint on their occurrence rate. Finally, in \S\ref{sec:conclusion}, we summarize our conclusions.
 


\section{Datasets} \label{sec:dataset}

\subsection{The eROSITA catalog of X-ray-bright M dwarfs} \label{subsec:erosita}

Our starting point is the catalog of X-ray bright M dwarfs compiled by \citet{Magaudda2022} constructed by cross-matching two eROSITA surveys with a refined version of the SUPERBLINK catalog of bright, nearby M dwarfs ($J<10$ and $V-J>2.7$) by \cite{lepine2011}. Here, we give a brief summary of the main steps followed by \cite{Magaudda2022} to build this catalog. 
\citet{Magaudda2022} cross-matched the SUPERBLINK catalog by \cite{lepine2011} with Gaia DR2 \citep{gaia2018b}, incorporating Gaia-based distances from \citet{bailerjones2018}. Stars without Gaia photometry or distance information were excluded. Spectral types were determined using Gaia $G_{BP}-G_{RP}$ colors, following the main-sequence values from \citet{mamajek2013}\footnote{The table “A Modern Mean Dwarf Stellar Color and Effective Temperature Sequence” is maintained and updated by E. Mamajek: \url{http://www.pas.rochester.edu/~emamajek/EEM_dwarf_UBVIJHK_colors_Teff.txt}}. M giants were excluded based on their location in the Gaia color–magnitude diagram and the final catalog includes spectral types spanning K5 through M7. Late type M dwarfs are under-represented and not present unless they are particularly bright, due to the cuts $J<10$ and $V-J>2.7$. Stellar parameters were derived using empirical relations from \citet{Mann2015, Mann2016}, calibrated using spectroscopically confirmed M dwarfs.

This refined version of the SUPERBLINK catalog was then cross-matched with two eROSITA surveys in order to identify X-ray–bright dwarfs: (1) the Final Equatorial Depth Survey (eFEDS; \citealt{brunner2022}), covering 142 deg$^{2}$ in the southern hemisphere and (2) the first release of the all-sky survey  (eRASS1; \citealt{predehl2021}). The result of this cross-match is a catalog of $687$ X-ray–emitting M dwarfs.

Neither of these eROSITA surveys provides the temporal resolution required to observe flares that are theorized to last minutes to hours \citep{gudel2009, yan2021, seli2025}. Specifically, in the eROSITA all-sky survey, the exposure time is $\sim40$\,s, with a cadence of $\sim6$ visits per day. In the eFEDS field, each source is observed for up to $\sim4.7$\,min in the central part of the field, and for shorter durations in the peripheral scans, which is still only marginally sufficient to detect most flares.

\subsection{The Chandra Sample}

\subsubsection{Parent Sample Selection}
\begin{figure*}
    \centering
    \includegraphics[width=8cm]{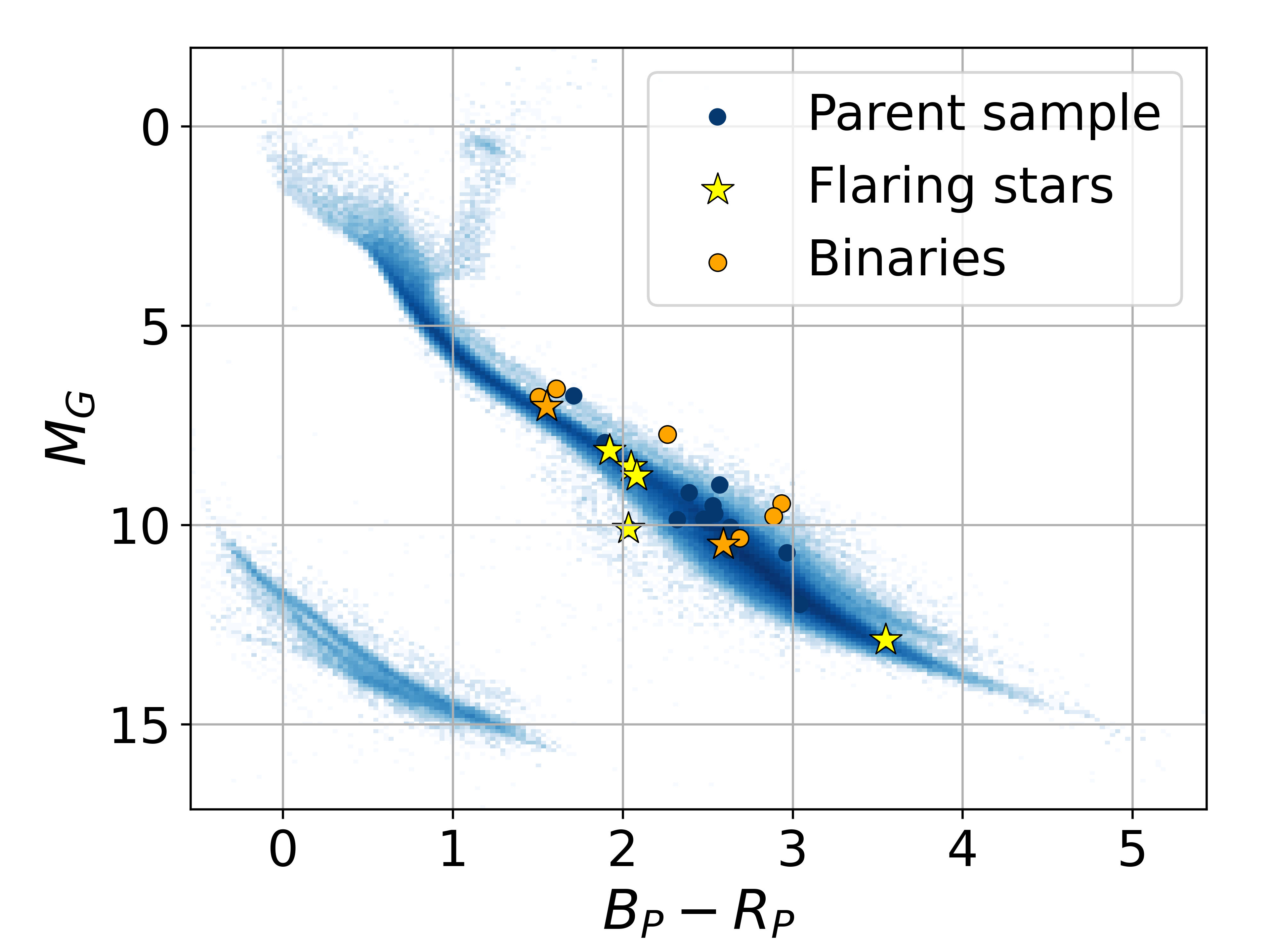}
    \includegraphics[width=8cm]{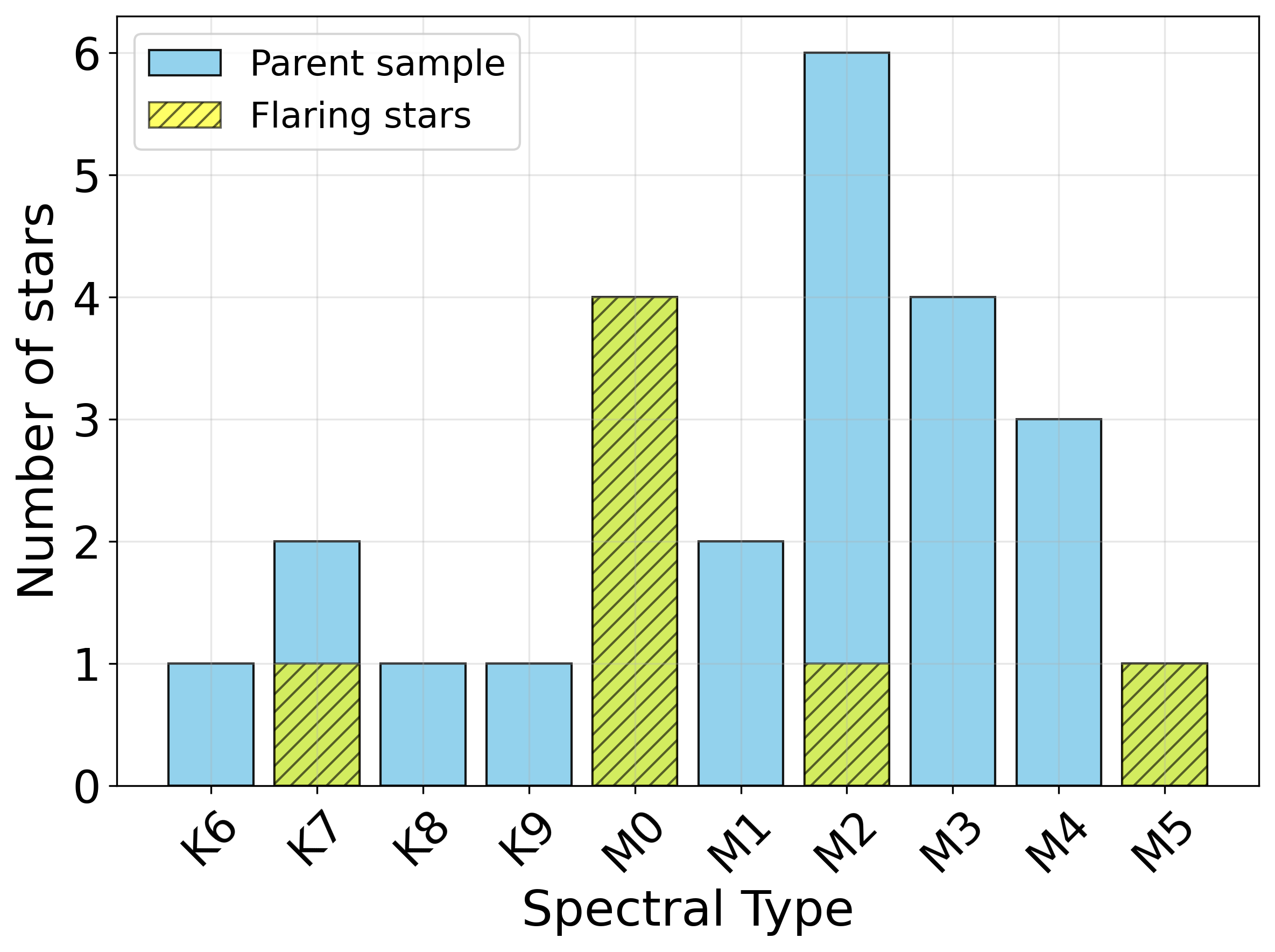}
    \includegraphics[width=8cm]{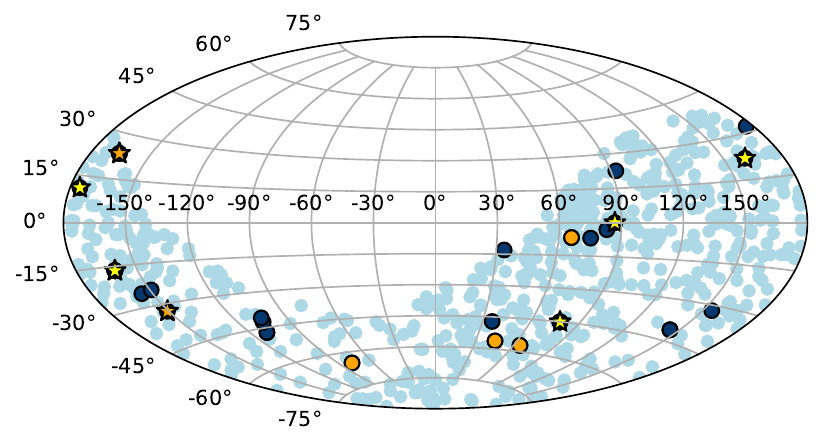}
     \includegraphics[width=8cm]{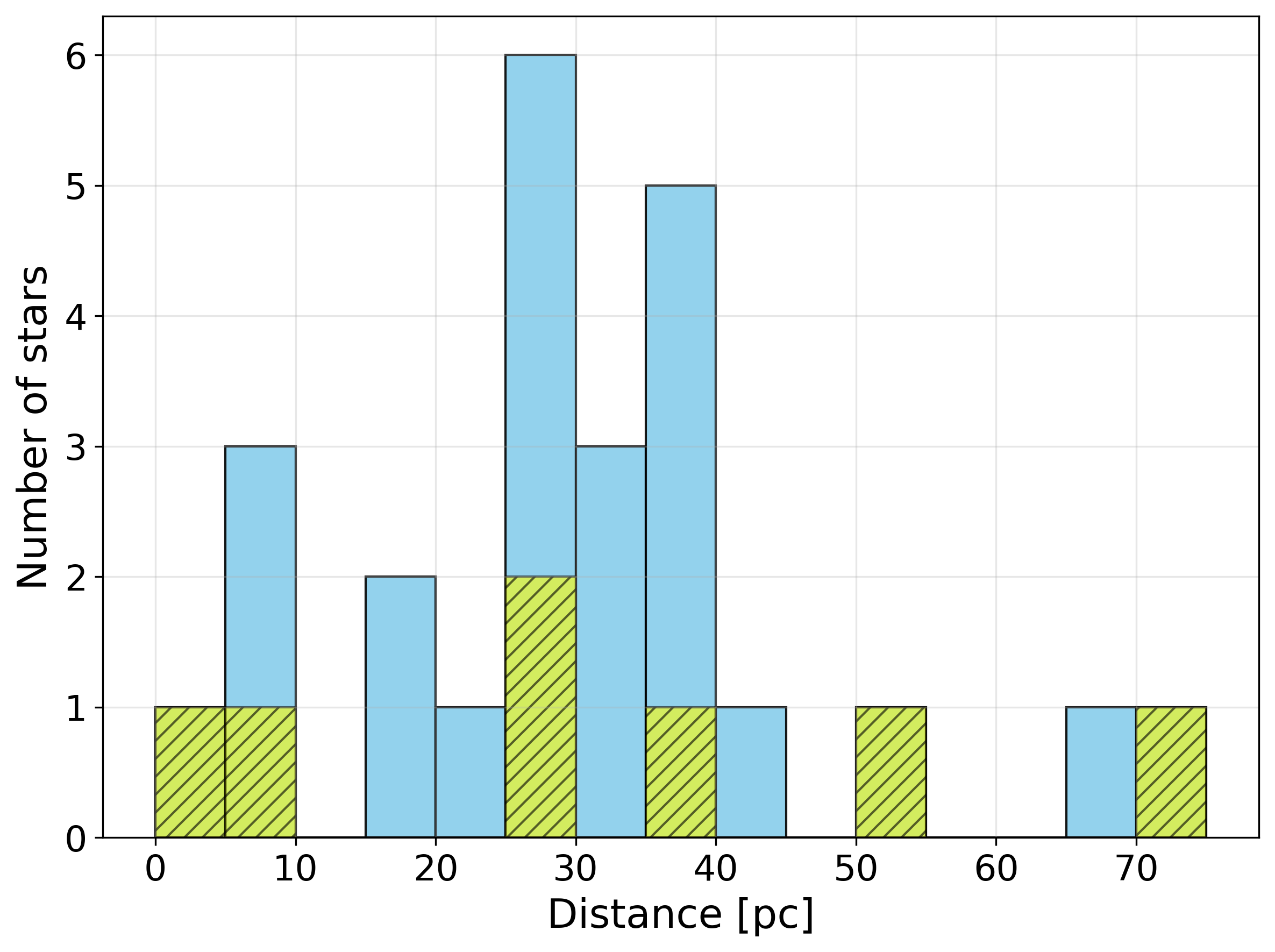}
    \caption{
{\bf Parent sample: M dwarfs presents in both eROSITA and the Chandra archive.}{\em Upper left:} Gaia DR3 color--magnitude diagram (CMD) constructed from BP--RP color and absolute $G$ magnitude \citep{gaia2023}. Light--blue points show Gaia DR3 sources within 100 pc (selected following \citealt{babusiaux2018}). Blue points indicate the stars in our sample, Orange points mark either known binaries or high-RUWE (likely binary) stars and yellow star symbol mark the subset of M dwarfs for which we detected flares. 
{\em Upper right:} Histogram of the spectral types for the parent sample (light blue) and for the stars in which flares were observed (yellow). Known binaries are excluded from this histogram. 
{\em Lower right:} Distance distribution of the parent sample (light blue) and the subsample for which we detect flares (yellow). 
{\em Lower left:} Aitoff projection in Galactic coordinates showing the sky distribution of the parent eROSITA catalog (light blue), the stars in the parent sample (blue), and the stars we identified as flaring(yellow).
}\label{fig:sample_graph}
\end{figure*}

We constructed a sample of stars present in both eROSITA and Chandra observations, by cross-matching the sample by \cite{Magaudda2022} described in the previous section with the Chandra archive \citep{weisskopf2000}. 
Proper motions were propagated to the epochs of the Chandra pointings to ensure accurate associations. We kept only normal Time Exposure (TE) mode ACIS observations. 
This resulted in a sample of $25$ distinct stars observed across $40$ different Chandra observations and during a cumulative exposure time of $881\rm\,ks$. 
We checked the Renormalized Unit Weight Error (RUWE) parameter\footnote{“Renormalizing the astrometric chi-square in Gaia DR2”; see: \url{https://www.cosmos.esa.int/web/gaia/public-dpac-documents}} from Gaia DR3 \citep{gaia2023}. Stars with $\mathrm{RUWE}>1.4$ values were inspected individually and identified as binaries, either through their SIMBAD classification \citep{wenger2000}, their presence in the Washington Double Star Catalog (WDS; \citealt{mason2001}), or their location in the color–magnitude diagram (see Figure~\ref{fig:sample_graph}). 


In Table~\ref{table_per_star}, we list the Gaia source ID, Chandra source ID, Chandra observation ID, distance, spectral type, stellar parameters, and RUWE value for each M dwarf in our sample, outlining in bold the cases for which one or more flares were identified. 
Figure~1 summarizes the properties of the objects in our parent sample, which include spectral types from late-K to mid-M dwarfs and distances out to 75~pc.



\subsubsection{Source Detection with the \texttt{X-Sifter} Pipeline} \label{xsifter}

We ran the {\tt X-sifter} source-detection pipeline \citep{soumagnac2024} on the $40$ Chandra observations covering the $25$ stars of our parent sample. 
Only photons in the $0.3$–$7$keV energy range were analyzed, following the prescription of \citet{evans2024} to mitigate calibration issues while retaining a substantial fraction of the photons detected near the lower-energy bound. All objects observed during the time span of the Chandra Source Catalog (CSC)~2.1 \citep{evans2024} (i.e., until 2022) have a counterpart in CSC~2.1 within $<1.3''$, except one. This is consistent with this object being detected by {\tt X-sifter} close to the detection limit, where the reported sensitivity of {\tt X-sifter} is approximately $1.3\times$ higher than that of the CSC, corresponding to a $\sim30\%$ increase in the number of detected (real) sources.

{\tt X-sifter} provides aperture photometry and light curves for the detected sources. In Figure~\ref{fig:combined}, we show the light curves of all the flares reported in this paper.

\subsection{Literature Sample}\label{sec:lit}
We conducted a literature review and compiled a list of the X-ray flares from M dwarfs which have been reported to date. The compiled sample is summarized in Table~\ref{table_per_flare} (per flare) and Table~\ref{table_per_star} (per star), together with key properties. Our review combines results from multiple references spanning the period from $2000$ to the present, and includes a total of $122$ flares from  $15$ distinct M dwarfs, observed during a cumulative duration higher than $1401.76\,\rm ks$ (this lower limit is the sum of all the observations for which the authors provided exposure times).

\begin{figure}
    \includegraphics[width=8cm]{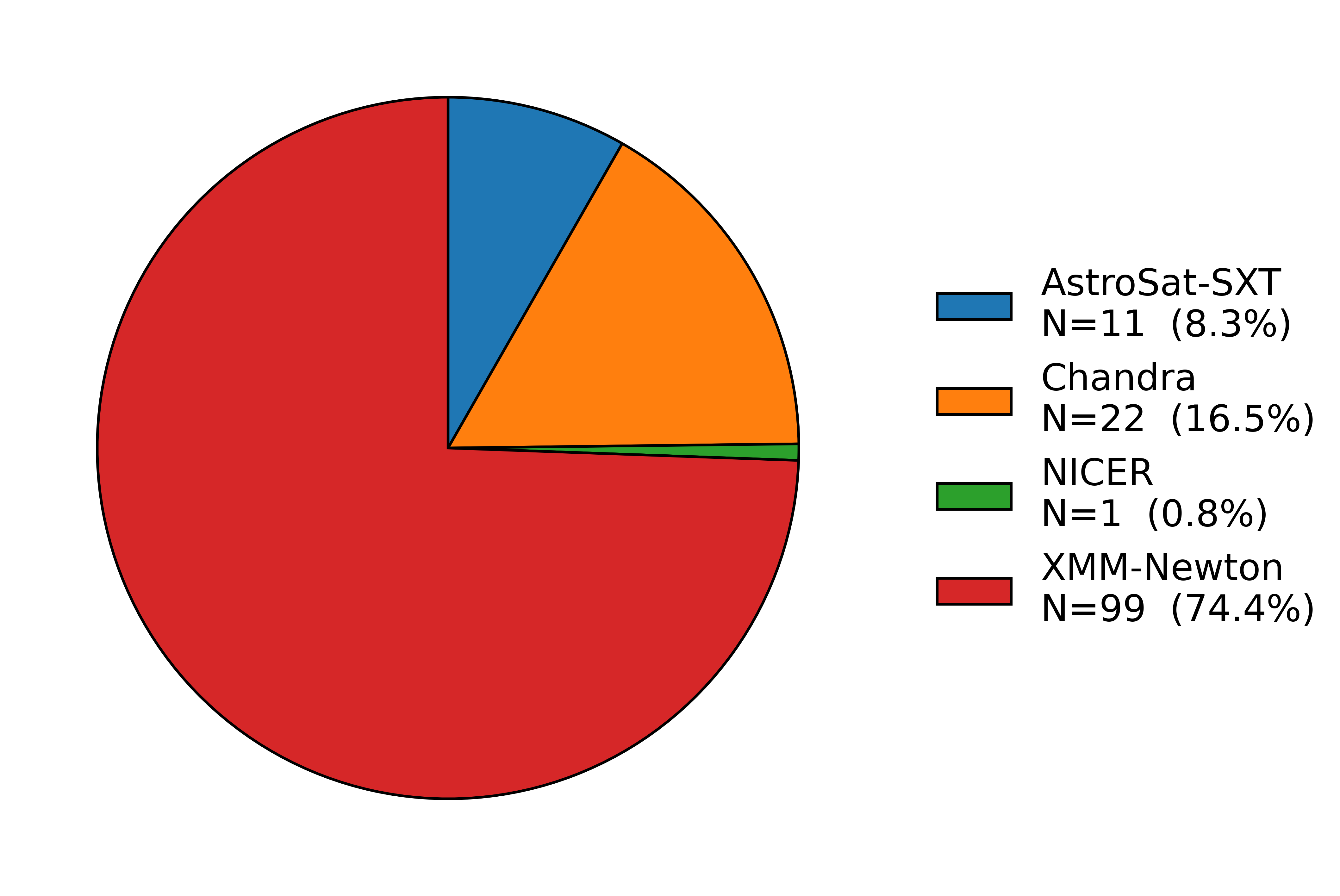}
    \caption{The instruments which captured flares from M dwarfs in the X-ray up to the publication date of this work.} 
    \label{fig:pie_exposure}
\end{figure}

The flares in the literature sample were discovered with a variety of X-ray telescopes, as illustrated in Figure~\ref{fig:pie_exposure}. Notably, the vast majority of X-ray flares where discovered with XMM-Newton.
A substantial fraction ($23\%$) of the flares in this sample were reported by \citet{pye2015}, who conducted the first blind, large-scale survey of stellar X-ray flares, based on a cross-match between the XMM-Newton Serendipitous Source Catalog \citep{watson2009} and the Hipparcos–Tycho stellar catalog \citep{hog2000}. They identified thirty one flares from nine M dwarfs. Another $\sim 40\%$ of the flares in the literature sample originate from an intensive observation campaign of a single star using XMM-Newton, AU\,Mic, by \cite{tristan2023}.
Table~\ref{table_per_star} shows a summary of the time spent observing the stars in both our sample and the literature sample.

We note that the literature sample is heterogeneous by construction: the reported flare properties originate from a wide variety of instruments and references, with a strong heterogeneity in the amount of information provided by each reference. The properties, when available, are used as they were reported, and we do not attempt to re-derive them.

\subsection{LP\,94420}\label{sec:LP}
We added one more object to this study. LP\,94420, the first brown dwarf and the first ultracool M-type object ever reported to be flaring in the X-ray by \cite{rutledge2000}. It was detected by Chandra but is not included in the SUPERBLINK catalog, since it has an observed magnitude $J=10.725$ and the SUPERBLINK catalog only includes objects with $J<10$. We ran {\tt X-sifter} on the relevant observation ({\tt OBSID}$=624$), detected the flare discovered by \cite{rutledge2000} and made our own derivation of the parameters explored in this study. This is the only case where we re-derived the parameters of a flare from the literature sample.

\section{Flares Detection} \label{sec:flares_identifications}

Below, we present our methodology for detecting flares in the light curves produced by {\tt X-sifter} and shown in Figure~\ref{fig:combined}. 

\subsection{Binning the Light Curves}\label{sec:binning}

Because X-ray observations are photon-starved, the choice of time binning can strongly affect our ability to detect flares. 
The challenge is to select a bin size that is large enough to contain sufficient counts, while remaining short enough to capture short-timescale variability in the light curve. To meet this requirement, we employ the Bayesian Blocks algorithm \citep{Scargle2013}, which partitions unbinned time-series data into segments (or ``blocks'') with statistically consistent count rates. 
When multiple blocks are identified, we subdivide the smallest block into \(a\) equal-length sub-blocks and require that each sub-block contains at least \(n_{\rm thresh}\) photons. 
If this criterion is satisfied, we bin the entire light curve using a bin width equal to \(1/a\) of the duration of the smallest block. If any sub-block falls below the \(n_{\rm thresh}\) threshold, we incrementally increase the bin size until all sub-blocks meet the criterion. In this work, we adopt $n_{\rm thresh}=3$ and $a=5$.
The resulting bin sizes are similar to those commonly adopted in other X-ray studies (e.g., \citealt{pye2015, Kuznetsov2021}), typically of order a few hundred seconds. 
In most cases, the flare corresponds to the smallest Bayesian block. .

\subsection{Flare Detection in the Binned Light Curve} \label{methodology}

To identify flares in the binned light curves, we adopt the algorithm  by \citet{Rekhi2023}, originally developed for the near-ultraviolet (NUV). 
The approach of \citet{Rekhi2023} is based on an adaptation of the FLAIIL pipeline introduced by \citet{Loyd2018b}. We refer the reader to \S\,1.3 and Figure\,5 of \citet{Rekhi2023} for a full description of the algorithm. 
 In Table~\ref{table_per_flare}, we list the SNR  obtained for each flare in our sample.


\section{Flares properties} 
\label{sec:flares_characterization}

In this section, we present the derivation of the various flares properties in the combined sample (this work+literature). The mean values, median values and uncertainties on these properties are presented in Table~\ref{tab:parameters_new}.

\begin{deluxetable}{lcccccc}
\tablecaption{{\bf Mean and median flare parameters.}
$t_{\rm flare}$, $t_{\rm rise}$, and $t_{\rm fall}$ denote the total flare duration, rise time, and decay time, respectively. $E$ is the flare energy. $F_f$ and $T_f$ are the best-fit flare flux and temperature, while $F_f/F_q$ and $T_f/T_q$ are the ratios of flare to quiescent flux and temperature. Spectral fits were performed using \texttt{UltraNest} \citep{buchner2021} and {\tt XSPEC} \citep{arnaud1996}.}

\label{tab:parameters_new}

\tablehead{
\colhead{Parameter} &
\multicolumn{3}{c}{This work} &
\multicolumn{3}{c}{This work + Literature} \\
\cline{2-4} \cline{5-7}
& \colhead{Mean} & \colhead{Median} & \colhead{$1\sigma \ interval$}
& \colhead{Mean} & \colhead{Median} & \colhead{$1\sigma \ interval$}
}

\startdata
$t_{flare}$ [s]    & 2164.05 & 1756.00 & [1034.8,3170.4] & 3695.40 & 2280.00 & [1322.00,5716.80] \\
$t_{rise}$ [s]   & 1080.56 & 996.90 & [483.77,1577.69] & 914.46 & 790.0 & [353.20, 1418.40] \\
$t_{fall}$ [s]   & 1198.1 & 1094.7 & [326.44,1815.72] & 1532.64 & 1080.00 & [622.75,2420.80] \\
E [$10^{31}$ erg] & 59.318 & 8.669 & [0.562, 19.420] & 39.659 & 8.854 & [1.953,48.799] \\
$F_{f}$ [$10^{-13} erg\, s^{-1}\, cm^{-2}$]    & 9.224 & 5.595 & [2.875,12.540] & \nodata & \nodata &\nodata \\
$F_f/F_q$  & 5.111 & 3.223 & [2.412,5.352] & \nodata & \nodata &\nodata \\
$T_{f}$  [keV] & 1.827 & 1.462 & [1.089,2.232] & \nodata & \nodata &\nodata \\
$T_f/T_q$   & 2.081 & 1.777 & [1.159,2.861] & \nodata & \nodata &\nodata \\
\enddata

\end{deluxetable}

\subsection{Light curves \& hardness ratio}\label{sec:hardness_ratio}

In Figure~\ref{fig:combined}, we show the light curves of the $11$ flares from the eROSITA–Chandra cross-match, as well as the flare of LP\,944-20 (see \S~\ref{sec:LP}). Specifically, we show the temporal evolution of (1) the photon counts, (2) the binned total photon energy, (3) the individual photon energies, and (4) the hardness ratios HR1 and HR2, defined following \citet{pandey2008} as the medium-to-soft and hard-to-medium energy band ratios, respectively.

Variations in the hardness ratios (HR) are predicted to trace flares, by models in which magnetic energy release heats plasma to high temperatures, producing harder X-ray emission \citep{reale2007}. This behavior has been observed in X-rays by \citet{pye2015} and \citet{Robrade2005}. We observe good agreement between the light curves and the temporal behavior of the HRs in most cases. 
We find good agreement between the light curves and the temporal evolution of the hardness ratios for {\tt ObsID}s 21071, 27611, 16302, and 24991 ($r_{\rm Pearson}>0.5$). Moderate agreement is observed for {\tt ObsID}s 17196 and 20138 ($r_{\rm Pearson}>0.2$), while no clear agreement is found for {\tt ObsID}s 17654 and 25686, as shown in Figure~\ref{fig:combined}. A more precise spectroscopic analysis is presented in the next sub-section. 

\subsection{Flux\textit{} \& Temperature}\label{sec:spectral}

For each observation containing identified flares, we performed the spectral analysis separately for the flaring phase and for the quiescent phases before and after the flare (except in the case of {\tt ObsID}$=20138$, as the observation begins during the detected flare). 
The spectra were extracted using the \texttt{specextract} tool from {\tt CIAO} version 4.17 \citep{ciao2006}, and spectral fitting was performed with {\tt XSPEC}, version 12.14.1 \citep{arnaud1996}.
To model the flare temperature, we adopted a single-temperature \texttt{apec} model \citep{smith2001}. Following \citet{Magaudda2022}, we fixed the abundance to $Z = 0.3,Z_{\odot}$, a typical value for coronal abundances in late-type stars \citep{favata2000,vandenbesselaar2003,Robrade2005,maggio2007}. The plasma temperature ($kT$) and normalization were left as free parameters, with parameter ranges set to $kT \in [0.1, 10.0]$~keV and normalization spanning $10^{-6}$ to $10^{-1}$, with log-uniform priors whose range was chosen as follows. The temperature prior spanned $kT \in [0.1, 10]$ keV, corresponding to plasma temperatures of approximately $1$--$120$ MK, encompassing both quiescent and flaring coronal temperatures reported for active M dwarfs (e.g., \citealt{gudel2004,reale2007}).

The APEC normalization, defined as
\begin{equation}
\mathrm{Norm}
=
\frac{10^{-14}}{4\pi D^{2}}
\int n_{\mathrm{e}}\, n_{\mathrm{H}} \, dV ,
\end{equation}

(where $D$ is the distance to the source (in cm), $n_{\mathrm{e}}$ is the electron number density, and $n_{\mathrm{H}}$ is the hydrogen number density) was assigned a log-uniform prior [$10^{-6}$--$10^{-1}$] (based on emission measures reported in \citealt{gudel2004,kowalski2024}).



\begin{figure*}
    \centering
    \includegraphics[width=8cm]{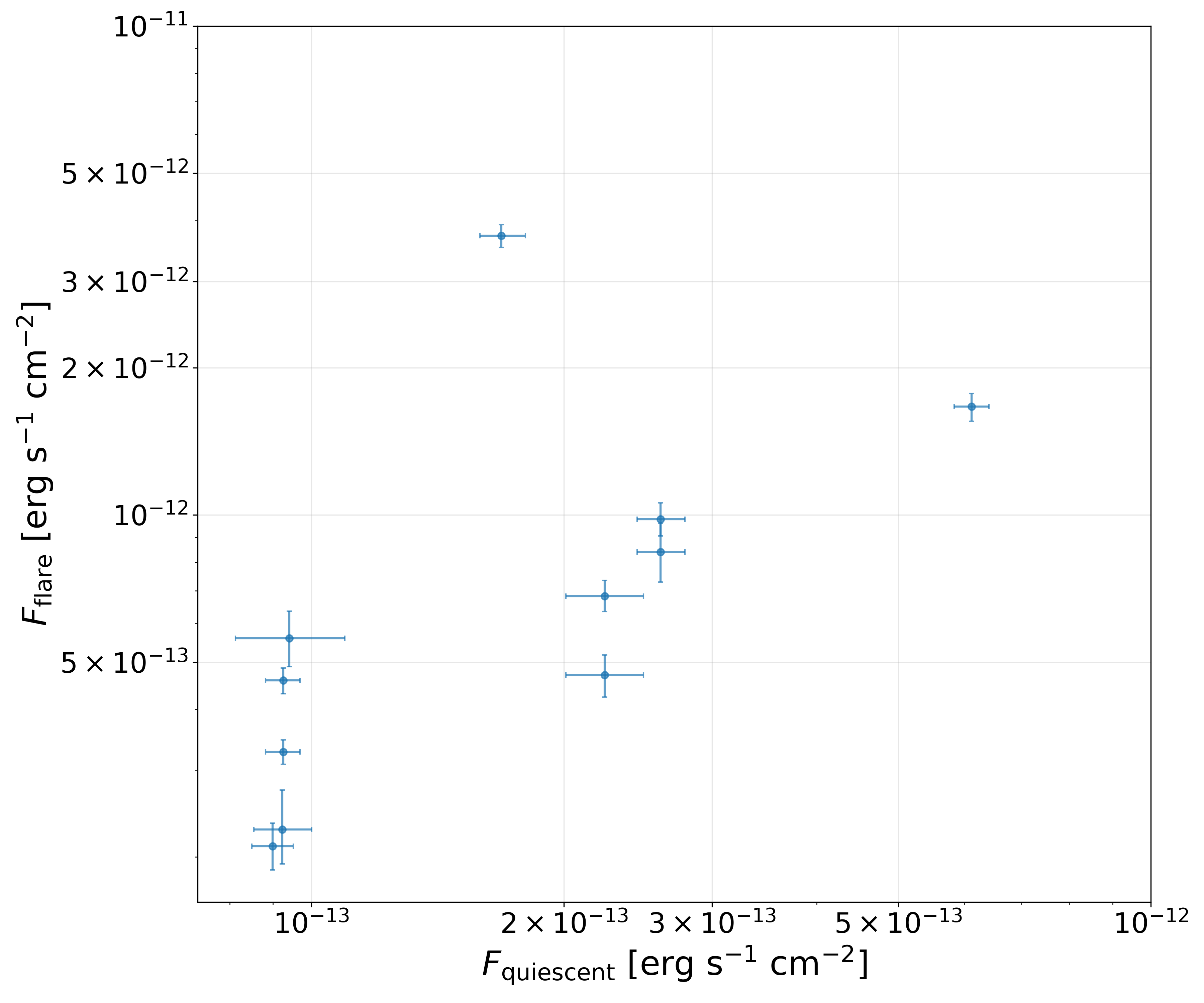}
    \includegraphics[width=8cm]{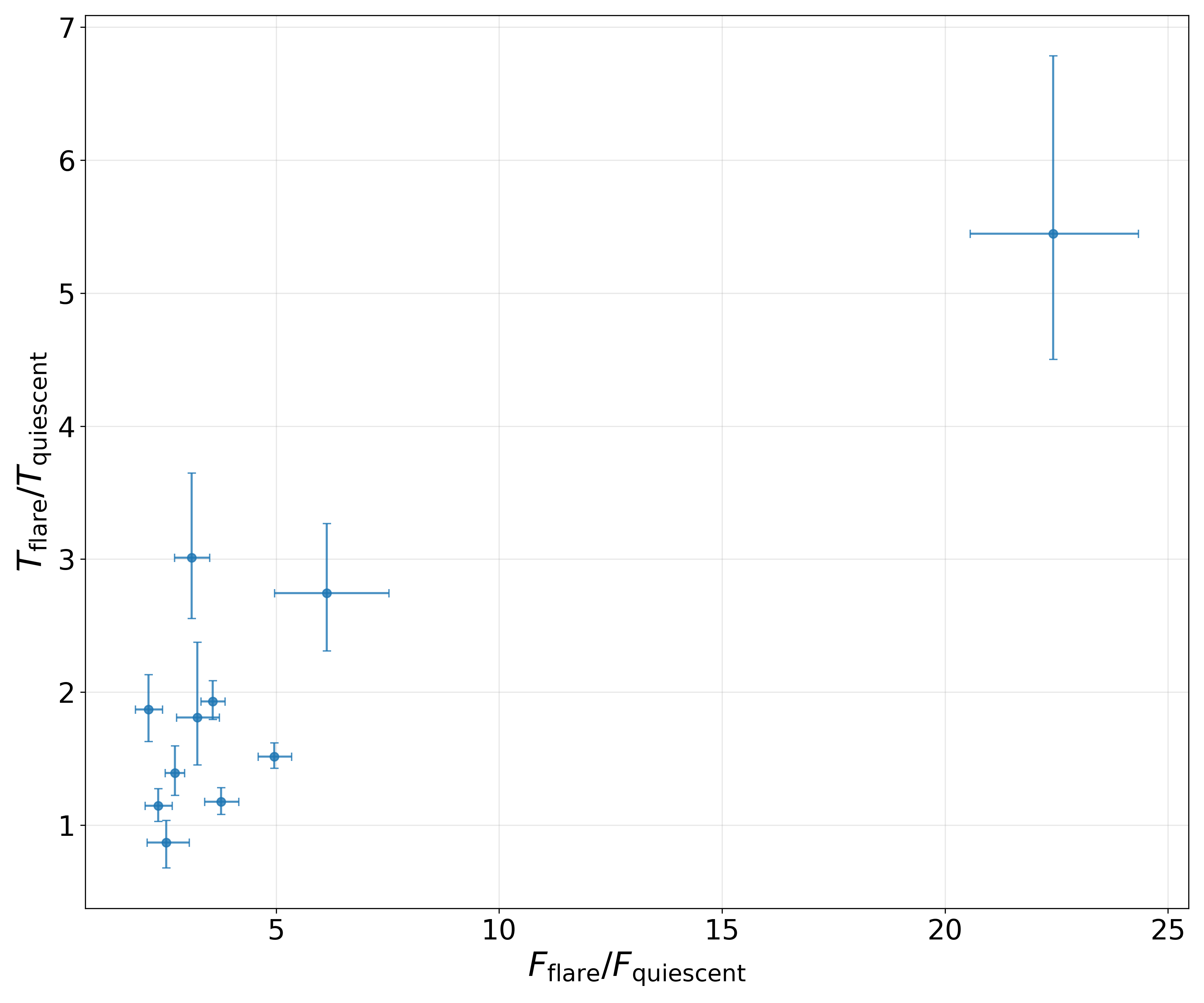}
    \caption{Left: The Flux measured during the flare, as a function of the flux measured during the quiescent phase (estimated as the minimum of the fluxes measured before and after the flare), for the $11$ flares we report. Right: The temperature increase during the flare (defined as the ratio of the temperature measured during the flare to the temperature measured during the quiescent phase), as a function of the flux increase (defined in the same way)}
    \label{fig:flux_flare_quiescent}
\end{figure*}

Best-fit parameters were determined using nested sampling via the \texttt{UltraNest} Python package \citep{buchner2021}, which implements the MLFriends algorithm \citep{buchner2016,buchner2019} to minimize the fit statistic—in this case, the C-statistic \citep{cash1979}, related to the log-likelihood through $C(\theta) = -2\ln \mathcal{L}(\theta)$. The best-fit parameters and their $1\sigma$ uncertainties are reported in Table~\ref{table_per_flare}. 
In Figure~\ref{fig:flux_flare_quiescent} we show the quiescent flux as a function of the flare flux.

We treat the temperature estimates with caution, as they can be underestimated due to a temperature–normalization degeneracy that arises in the low–photon-count regime ($\lesssim 50$ counts). This degeneracy, and its biasing effect on temperature estimates in low-photon X-ray spectra, is well documented \citep{nousek1989,loredo1992,park2006}, including within the {\tt XSPEC} fitting methodology \citep{arnaud1996}.

Fluxes in the 0.3–7.0~keV energy band were computed using {\tt XSPEC} \citep{arnaud1996}, applied to the posterior distribution from the {\tt UltraNest} fit. The resulting values are presented in Table~\ref{table_per_flare} and Table~\ref{tab:energy_table}. We note that, while the temperature–normalization degeneracy can bias temperature estimates in low-count spectra, the inferred X-ray flux is relatively robust, because changes in temperature are compensated by the normalization along the degeneracy, and fluxes are derived by marginalizing over the full posterior distribution.

The evolution of the temperatures and fluxes during the flares is shown in Figure~\ref{fig:temp_flux}. We measure a mean temperature of $1.8~ \rm KeV$, temperature increases of $0.9$ to $5.5$, a mean flux of $9.2~\times 10^{-13} \rm erg/s/cm^2$, and flux increases of $2.1$ to $22.1$ during the flare, as summarized in Table~\ref{tab:parameters_new}.  
These values are consistent with theoretical predictions by \citet{reale2007}, who predicted an increase to tens of millions of Kelvins during flares.


\subsection{Flare duration, rise time \& fall time}\label{sec:duration}
 
To characterize the duration and temporal shape of the flares, we adopt the following definitions: (1) the {\bf flare duration $t_{\rm flare}$} is the time span between the centers of the first and last time bins of the flare. The uncertainty on $t_{flare}$ is estimated as half of the adopted time-bin width. Since this estimator of the flare duration is likely to be biased (for example, when the flare starting time is buried below the background level), we also estimate the flare duration via the full-width at half maximum (FWHM) value of the flare. (2) The flare {\bf rise time $t_{\rm rise}$} is the interval between the center of the first time bin in which the flare is identified and the center of the time bin with the highest photon count.
(3) The {\bf decay time $t_{fall}$} is the interval between the center of the time bin with the highest photon count and the center of the first bin in which the flare is identified as having ended. (4) The {\bf flare asymmetry} is defined as the ratio of the rise time to the decay time.
To ensure reliable measurements, we restrict our analysis to flares that are observed in their entirety. For these events, the total flare duration is defined as the sum of the rise and decay times.

The measurements of all these properties are reported in Table~\ref{table_per_flare}.

\begin{figure*}
    \centering
    \includegraphics[width=8cm]{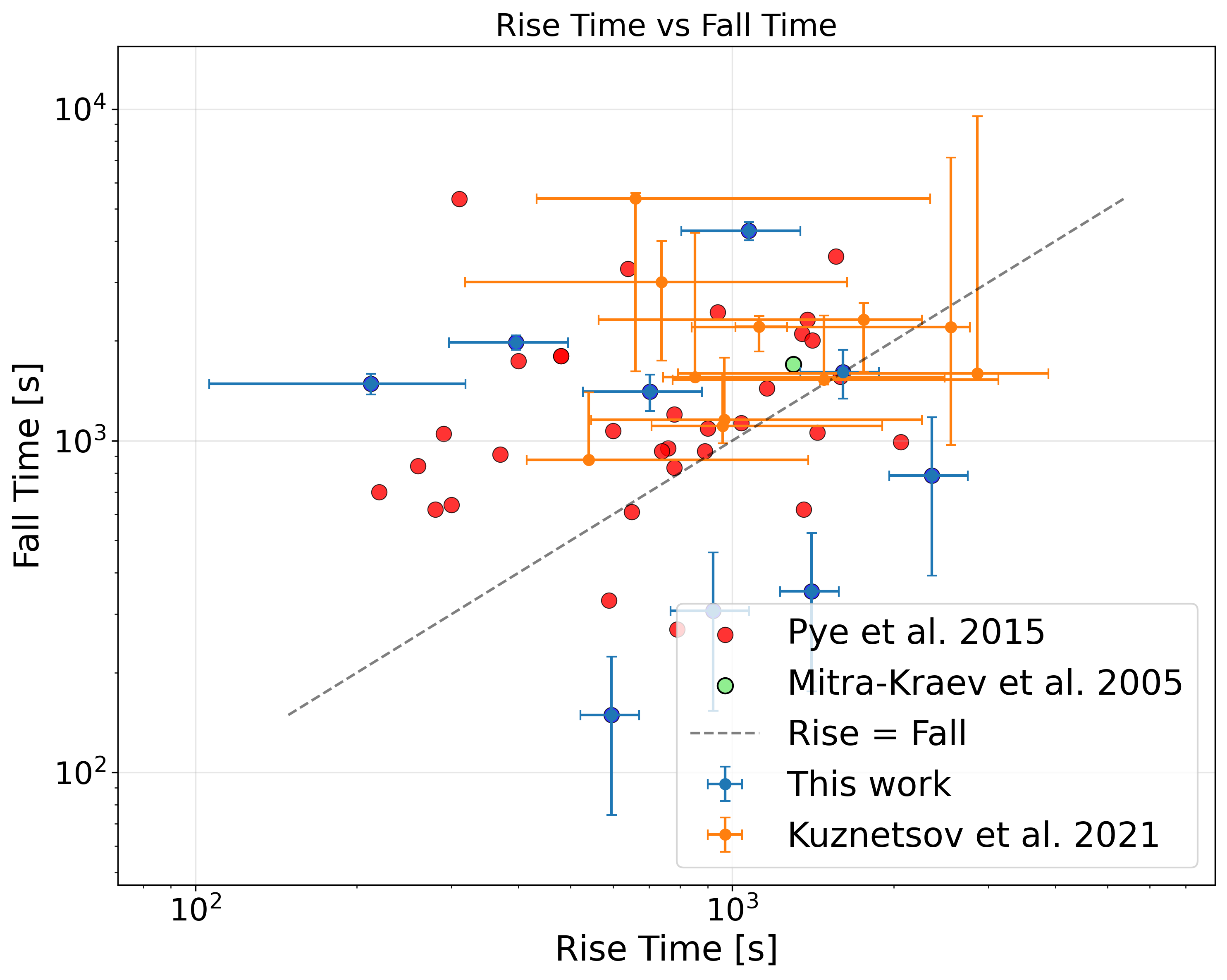}
    \includegraphics[width=8cm]{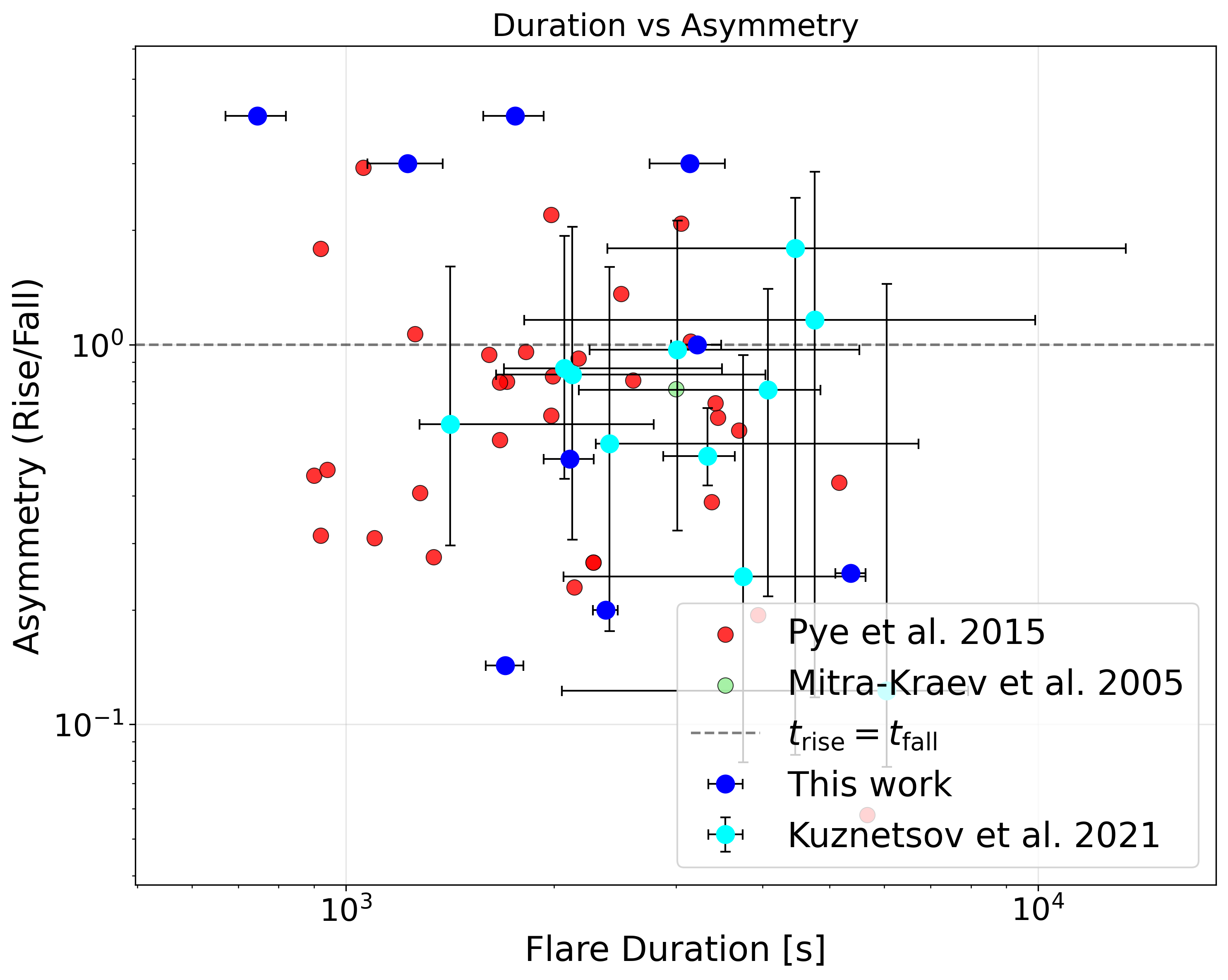}
    \caption{Left panel: A comparison between the flare rise and decay time for our sample (blue) and the literature sample (other colors). We include \cite{pye2015} but we note they did not provide error bars. Right panel: the asymmetry of the flare temporal shape as a function of the flare duration.}
    \label{rise_fall}
\end{figure*}

In Figures~\ref{rise_fall}, we show the flare rise times, decay times, and temporal asymmetry for the flares reported in this work, as well as for those previously reported in the literature (specifically by \citealt{Mitra2005}, \citet{pye2015}, and \citealt{Kuznetsov2021}). The durations of the $11$ flares reported in this work span values similar to those previously reported in the literature. For the full sample, we find a typical (mean) flare duration of $3.6~\rm ks$, as reported in Table~\ref{tab:parameters_new}. 

The classical picture of magnetic reconnection–driven flares predicts a rapid rise followed by a more gradual decay phase \citep[e.g.,][]{priest2002,kowalski2024}. 
To assess whether the current data support this picture, we performed a Bayesian model comparison between a model in which the rise and decay times are equal and a model that allows the decay time to differ from the rise time by a constant offset (which may be either positive or negative). We performed the model comparison using (1) only the flares reported in this work, and (2) the combination of the flares reported here and those reported in the literature (specifically by \citet{Kuznetsov2021}, as this is the only study that reports uncertainties on both rise and decay times). For events with asymmetric uncertainties, we adopted a conservative approach by using the larger of the upper and lower uncertainties when constructing the likelihood. 
Using \texttt{UltraNest}, we computed the Bayesian evidence for each model, which is reported in Table~\ref{tab:model_selection}. The model selection favors the asymmetric model in both cases, with a best-fit offset parameter of
$a = 342^{+64}_{-57}$ s for our sample alone ($a$ is defined as $t_{fall}=t_{rise}+a$), indicating that flare decay times are systematically longer than rise times.


\begin{table*}[t]
\centering
\caption{Bayesian Model Comparison of Flare Rise and Decay Times.}
\label{tab:model_selection}
\begin{tabular}{lcccccc}
\hline
Sample &
Model &
Relation &
$\log Z$ &
$\Delta \log Z$ &
Bayes Factor $K$ &
$a$ [s] \\
\hline
This work &
Symmetric &
$t_{\rm fall}=t_{\rm rise}$ &
$-203.6$ &
$0$ &
1 &
-- \\
 &
Offset &
$t_{\rm fall}=t_{\rm rise}+a$ &
$-190.5$ &
$+13.1$ &
$2.8\times10^{5}$ &
$342.20^{+64.2}_{-57.0}$ \\
\hline
This work + [1] &
Symmetric &
$t_{\rm fall}=t_{\rm rise}$ &
$-302.89$ &
$0$ &
1 &
-- \\
&
Offset &
$t_{\rm fall}=t_{\rm rise}+a$ &
$-287.58$ &
$+15.31$ &
$4.4\times10^{6}$ &
$366.65^{+58.8}_{-62.1}$ \\
\hline
\end{tabular}
\tablecomments{
Bayesian evidences were computed using nested sampling with \textsc{UltraNest}.
Bayes factors are defined as $K=\exp(\Delta\log Z)$.
Quoted uncertainties correspond to 68\% credible intervals. [1] is \cite{Kuznetsov2021}.
}
\end{table*}

\subsection{Flare Absolute Energy}

We compute the flare energy as
\begin{equation}\label{eq:energy}
E = 4\pi d^{2}\left(F_{\mathrm{flare}} - F_{\mathrm{quiescent}}\right) t_{\mathrm{flare}},
\end{equation}
where $d$ is the distance to the flaring star, $t_{\mathrm{flare}}$ is the flare duration, and
$F_{\mathrm{flare}}$ and $F_{\mathrm{quiescent}}$ are the measured fluxes during the flaring and quiescent phases, respectively.

To estimate the uncertainties on the energies, we used the  posterior distribution of fluxes computed in \S~\ref{sec:spectral}, and drew distances and durations from Gaussian distributions, using the uncertainties provided by \textit{Gaia} for the distances and the uncertainties adopted in \S~\ref{sec:duration} for the durations. 

In Table~\ref{tab:energy_table} of the Appendix~\ref{sec:tables},  we summarize the flare energies, as well as the quiescent and peak flare fluxes we derive. 
We note that we did not apply a correction for interstellar bound-free absorption to the observed flare fluxes. The stars in our sample are nearby and are therefore expected to lie within the Local Bubble, where typical hydrogen column densities are low, $\log N_{\rm H} \lesssim 19.3$ \citep{lallement2018,redfield2008,frisch2011}. Adopting $N_{\rm H}\approx2\times10^{19}\,\mathrm{cm^{-2}}$ and the interstellar absorption model of \citet{wilms2000}, we estimate an optical depth at $1\,\mathrm{keV}$ of only $\tau \approx 4\times10^{-3}$, corresponding to a flux attenuation of $\sim0.4\%$. This effect is negligible compared to the statistical and systematic uncertainties of our flare-energy estimates, and was therefore ignored throughout this work.

\begin{figure}
    \centering
    \includegraphics[width=8cm]{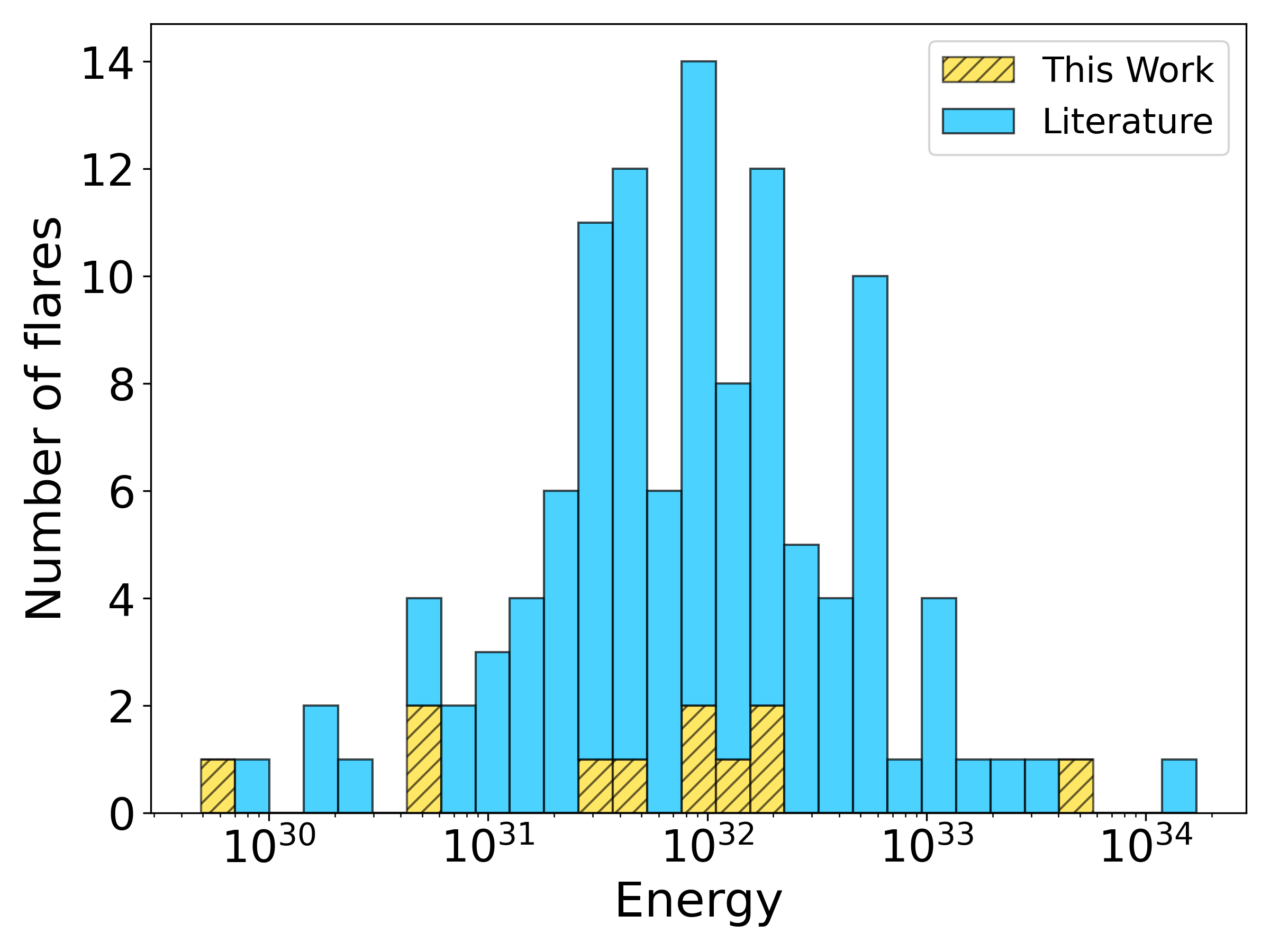}
    \caption{\textbf{Flare Energies Distribution} The blue bars represent the flares from the literature and the yellow flares are from this work. We note that $\sim 40 $ of the flares from the literature sample are from one single star.} 
    \label{fig:flare_energy_hist}
\end{figure}

Figure~\ref{fig:flare_energy_hist} shows the distribution of flare energies for the full sample, spanning four orders of magnitude, with a mean energy $E = 8.8 \times 10^{31},\rm erg$. 

\subsection{Correlation between Flare Duration and Flare Strength}\label{sec:energy-duration}
Long-duration flares are predicted by theoretical models to exhibit higher energies. Specifically, \citet{maehara2015} predicted a $t_{flare} \propto E^{1/3}$ dependence for solar-type star flares at optical wavelengths. The underlying idea is that larger magnetic structures produce flares that are both longer in duration and more energetic.

A power-law dependence has been observed in stellar flares at optical and UV wavelengths: \citet{howard2019} identified power-law duration–energy relations in \textit{Kepler} M-dwarf flares; \citet{tu2020} found similar relations for TESS superflares on solar-type stars; and \citet{yang2023} confirmed that these scaling relations persist in TESS optical flare of solar-type stars samples. For M dwarfs in particular, a power-law dependence has been observed at optical and UV wavelengths by \cite{howard2019,Yang2017}.

At X-ray wavelengths, \citet{veronig2002} observed a similar duration–energy correlation in the Sun. This result was later nuanced by \citet{reep2023}, who proposed that the duration–energy relation can be influenced by the physics of plasma cooling and by the heights at which the emission forms, potentially explaining wavelength-dependent differences in duration–energy behavior in solar-like stars. \citet{gudel2004} and \citet{kowalski2024} also proposed explanations for isolated cases in which long-duration flares are not particularly energetic, a behavior observed, for example, in flares on Proxima Centauri. Here, we aim to explore the energy–duration dependence in our combined sample.

\begin{figure*}
    \centering
    \includegraphics[width=8cm]{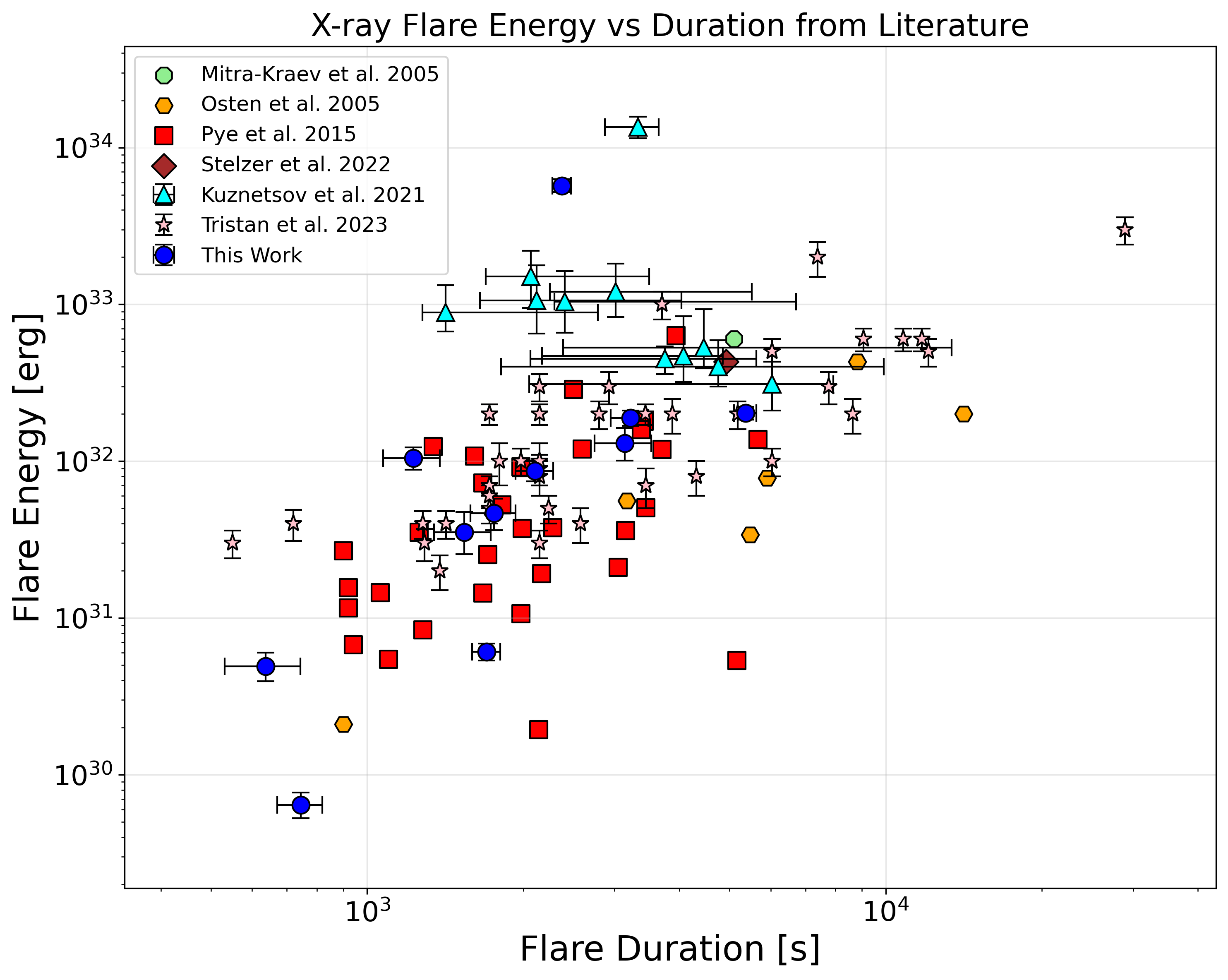}
     \includegraphics[width=8cm]
     {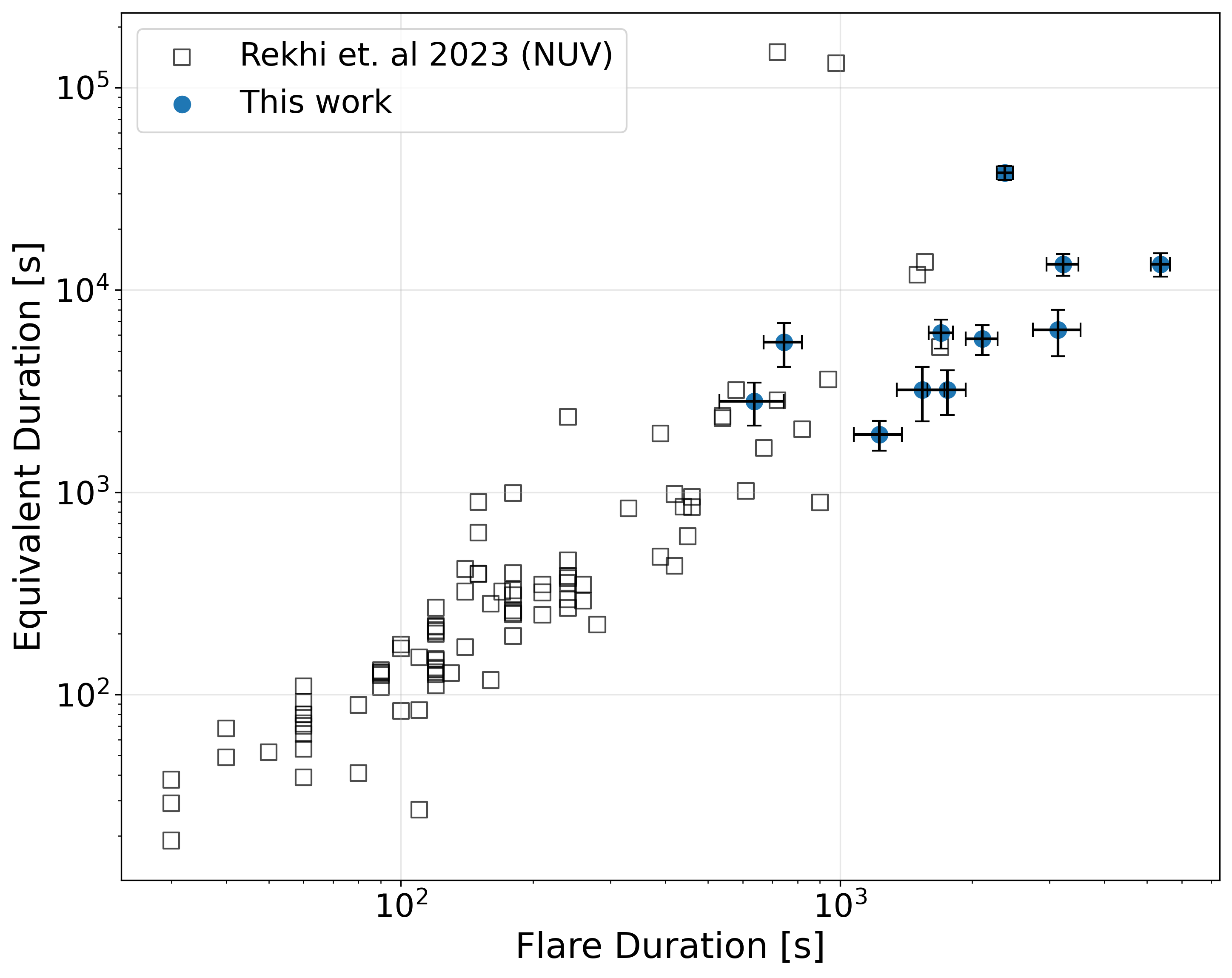}
    \caption{\textbf{The correlation between the flare duration and the flare strength.} Left panel: the flares energies as a function of their duration. Our sample is shown with blue dots. The literature data are plotted with different colors and markers. Right panel: the flares equivalent duration as a function of their duration. The black squares represent the NUV flares reported by \citet{Rekhi2023}. The blue circles correspond to the X-ray flares identified in this work. No equivalent duration was provided in the literature sample.}
    \label{fig:flare_energy_duration}
\end{figure*}

In Figure~\ref{fig:flare_energy_duration}, we show the energies measured for the combined sample as a function of the flare duration. 
For the data used in this work, we obtain a best-fit power-law index of $\alpha=0.34_{-0.01}^{+0.01}$, which we report in Table~\ref{tab:e_dur}. In this same table, we also report the results when using the FWHM of the light curve (in counts and in energy), a potentially less biased estimator of the duration.

 \begin{table}
 \centering \caption{Power-law Fit Parameters for the $t_{\rm flare}$-$E$ dependence.} 
  \label{tab:e_dur} 
\begin{tabular}{ccc} 
 \hline
  & $\log_{10} A$ & $\alpha$ \\ 
 \hline 
 $t_{flare}$ & $7.29_{-0.30}^{+0.31} $ & $0.34_{-0.00}^{+0.01} $ \\ 
 $FWHM_{counts}$ & $5.16_{-0.35}^{+0.36}$ & $0.27_{-0.01}^{+0.01}$ \\
 $FWHM_{Energy}$ & $4.42_{-0.33}^{+0.34}$ & $0.24_{-0.01}^{+0.01}$ \\ 
 \hline 
 \end{tabular}
 \end{table}


Another common way to quantify and compare flare strengths, independent of the temporal profile, duration, and absolute brightness of the flare, is the equivalent duration $\delta$, introduced by \citet{gershberg1972}, which, for light curves binned in discrete time intervals, can be written as

\begin{equation}
\delta = \sum_{i \in \mathrm{flare}} \frac{F_i - F_q}{F_q}  \Delta t_i\,,
\end{equation}
where $F_i$ is the observed count rate in the $i^{\rm th}$ time bin, $F_q$ is the quiescent count rate and $\Delta t_i$ is the width of the $i^{\rm th}$ bin. Physically, the equivalent duration represents the amount of time the star would need to emit the flare’s excess energy at its quiescent emission level. As such, it enables a comparison of flares that is independent of their absolute luminosities, temporal profiles, and durations.

In Table ~\ref{table_per_flare}, we show the values of $\delta$ calculated for the flares in our sample. Errors where computed using \cite{gehrels1986}. In Figure~\ref{fig:flare_energy_duration}, we show the flare equivalent duration as a function of total flare duration. We restrict this analysis to the flares reported in this work, as equivalent durations are not available for the literature sample. We overplot our measurements with those reported by \citet{Rekhi2023} for a sample of hundreds of NUV flares observed with \textit{GALEX}. Longer-duration flares tend to be associated with larger equivalent durations, consistent with the fact that longer events generally release more energy in these bands.

To quantify this trend, we fit a power-law relation between the equivalent duration and flare duration of the form
\begin{equation}
t_{\rm flare} = A\,\delta^{\alpha}.
\end{equation}
The best-fit parameters for the X-ray, NUV, and combined X-ray+NUV samples are reported in Table~\ref{tab:ed_dur}. All are consistent with the coefficient measured by \citet{Rekhi2023} at UV wavelengths.

\begin{table}
\centering \caption{Power-law Fit Parameters for the $t_{\rm flare}$-$\delta$ dependence.} 
\label{tab:ed_dur} 
\begin{tabular}{ccc} 
\hline
Band & $\log_{10} A$ & $\alpha$ \\ 
\hline 
X-ray & $1.57 \pm 0.74$ & $0.45 \pm 0.19$ \\ 
NUV & $0.99 \pm 0.07$ & $0.50 \pm 0.03$ \\
X-ray $+$ NUV & $0.83 \pm 0.07$ & $0.58 \pm 0.03$ \\ 
\hline 
\end{tabular} 
\end{table}

\section{Flaring stars properties \& Rates} \label{results}

\subsection{Spectral Types of Flaring M Dwarfs}
\begin{figure}
    \centering
    \includegraphics[width=8cm]{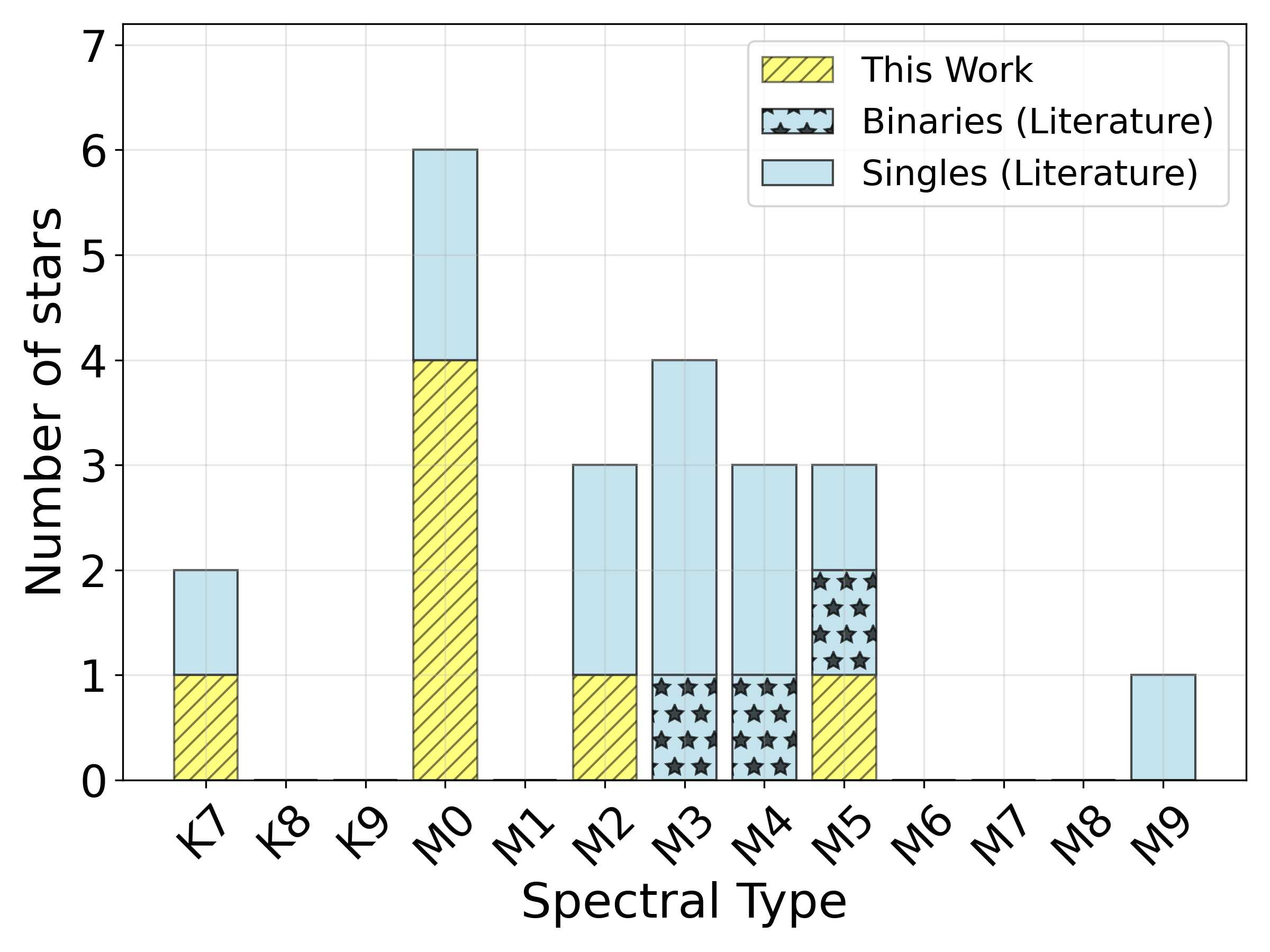}
    \caption{Spectral type distribution of the flaring stars identified in this work (yellow) and from the literature (blue).}
    \label{fig:hist_flare}
\end{figure}

Figure~\ref{fig:hist_flare} shows the distribution of identified flares as a function of spectral type within our sample, indicating that X-ray flaring activity is present across the full range of spectral types considered in this work. 

\subsection{Flare frequency}\label{sec:rates}

\subsubsection{Flaring time}
\begin{figure*}
        \includegraphics[width=9cm]{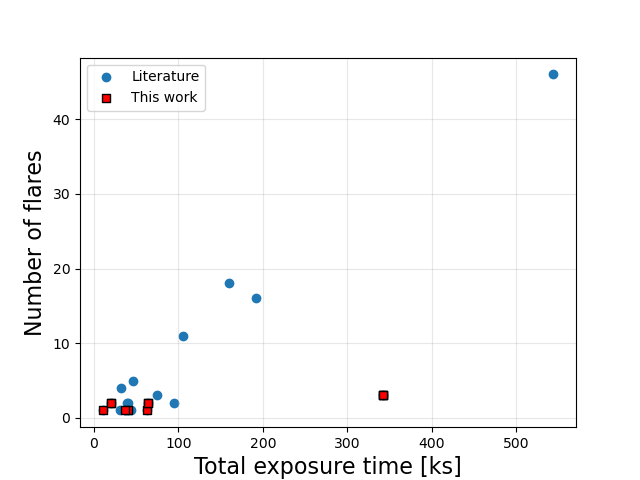}
        \includegraphics[width=9cm]{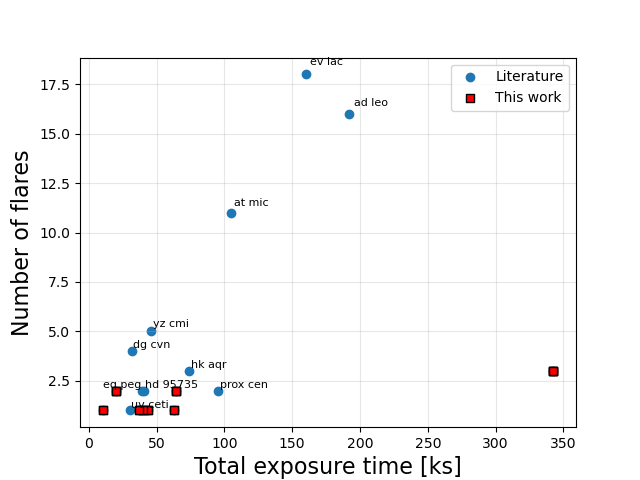}

    \caption{Left: The  number of observed flares and the exposure time, for all the stars in the full sample for which exposure times are available. Right: zoom-in view of the left panel.}
    \label{fig:hist_num_flares_exposure_literature}
\end{figure*}

\begin{figure}
        \includegraphics[width=9cm]{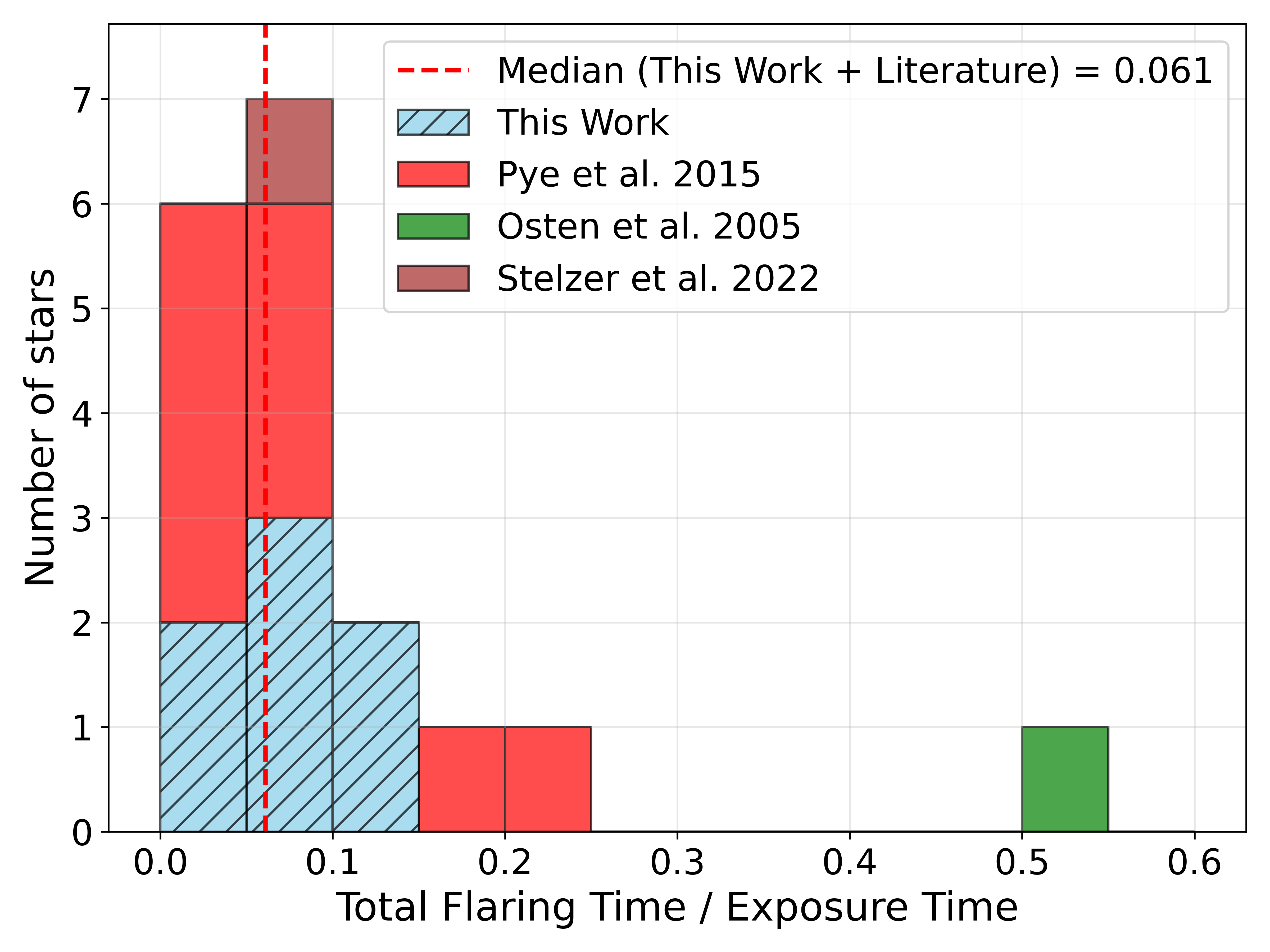}
\caption{Histogram of the percentage of observing time spent in a flaring state, for all the stars in the combined sample for which flare duration and observing time where provided.}
\label{fig:time-spent-flaring}
\end{figure}

In Figure~\ref{fig:hist_num_flares_exposure_literature}, we show the number of observed flares per star as a function of the total exposure time during which each star was observed 
In our combined sample (including flares reported in this work as well as in the literature), exposure times are available for $122$ flares, and the overall flare occurrence rate is $10^{-1}\,\mathrm{ks^{-1}}$. Figure~\ref{fig:time-spent-flaring} shows the relative time spent flaring per star: $50\%$ of the M dwarfs observed to flare in X-rays spend less than $\sim6\%$ of the total observed time in a flaring state.

\subsubsection{Flare Frequency Distribution (FFD) as a function of energy and spectral type}

\begin{table}
\centering
\caption{Power-law fit parameters for energy FFD}
\label{tab:ffd_energy}
\begin{tabular}{ccc}
\hline
Type & $\log_{10} C$ & $\alpha$ \\
\hline
M0--2  & $17.78^{+0.75}_{-0.67}$ & $0.54^{+0.02}_{-0.02}$ \\
M3--6  & $14.87^{+0.42}_{-0.45}$ & $0.45^{+0.01}_{-0.01}$ \\
\hline
\end{tabular}
\end{table}

In this section, we aim to estimate the rate of X-ray flares in M dwarfs, as a function of the flare energy and the M dwarf spectral type.

The occurrence rate of M dwarfs flares has been measured in a wide range of wavelengths and using chromospheric $H_{\alpha}$ variability (e.g. in SDSS;  \citealt{koller2021}). To our knowledge, this work is the first measurement of the flare occurrence rate at X-ray wavelengths.

A common way to quantify the occurrence rate is the cumulative Flare Frequency Distribution (FFD; the occurrence rate of flares with energies greater than a given energy $E$, also noted $\nu(E)$ in what follows). The FFD has been reported to follow a power law across a wide range of wavelengths, from the near-infrared to the UV \citep{Loyd2018,Murray2022,stelzer2022,Rekhi2023}.

\begin{figure*}
\includegraphics[width=9cm]{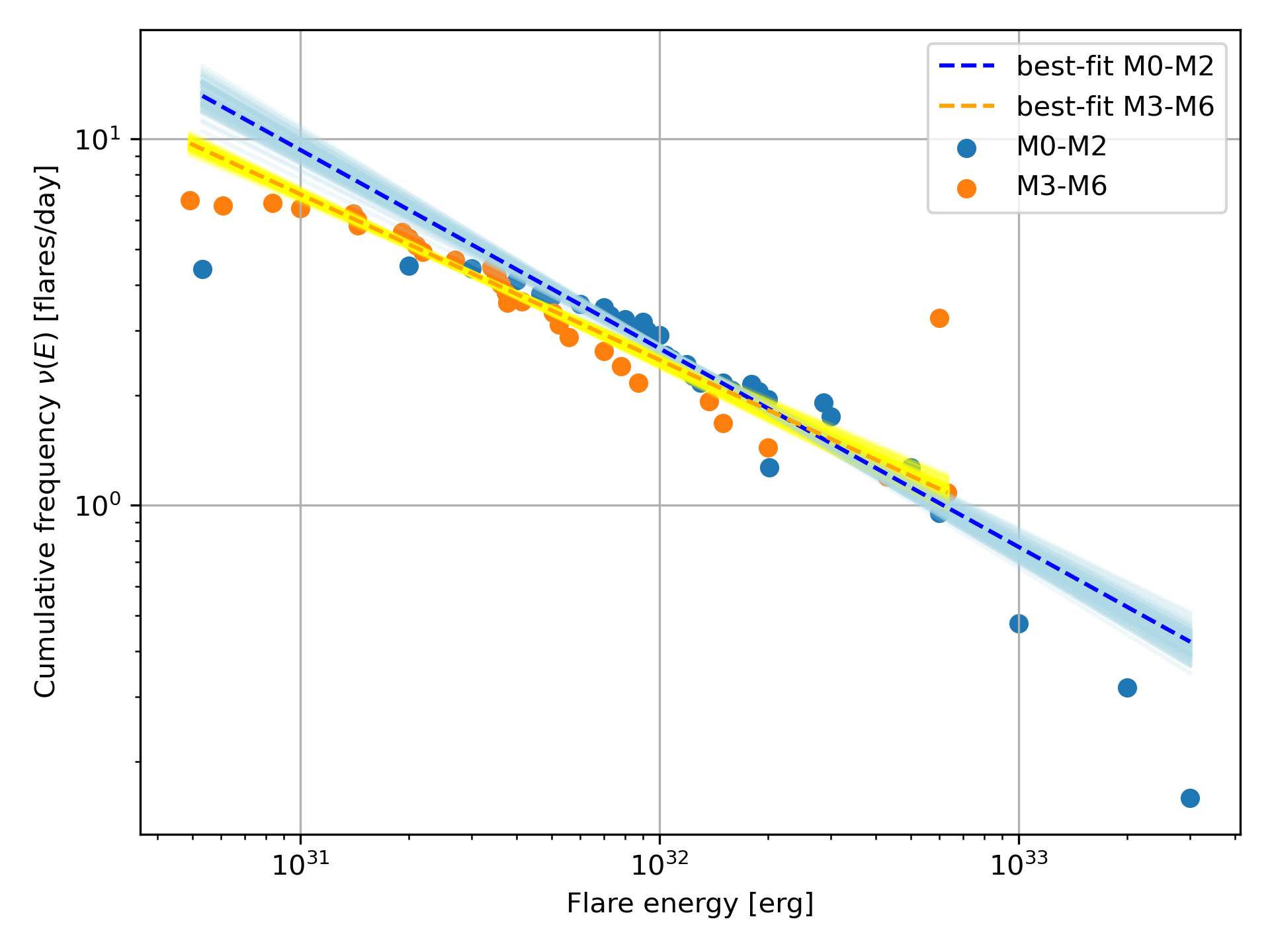}
\includegraphics[width=9cm]{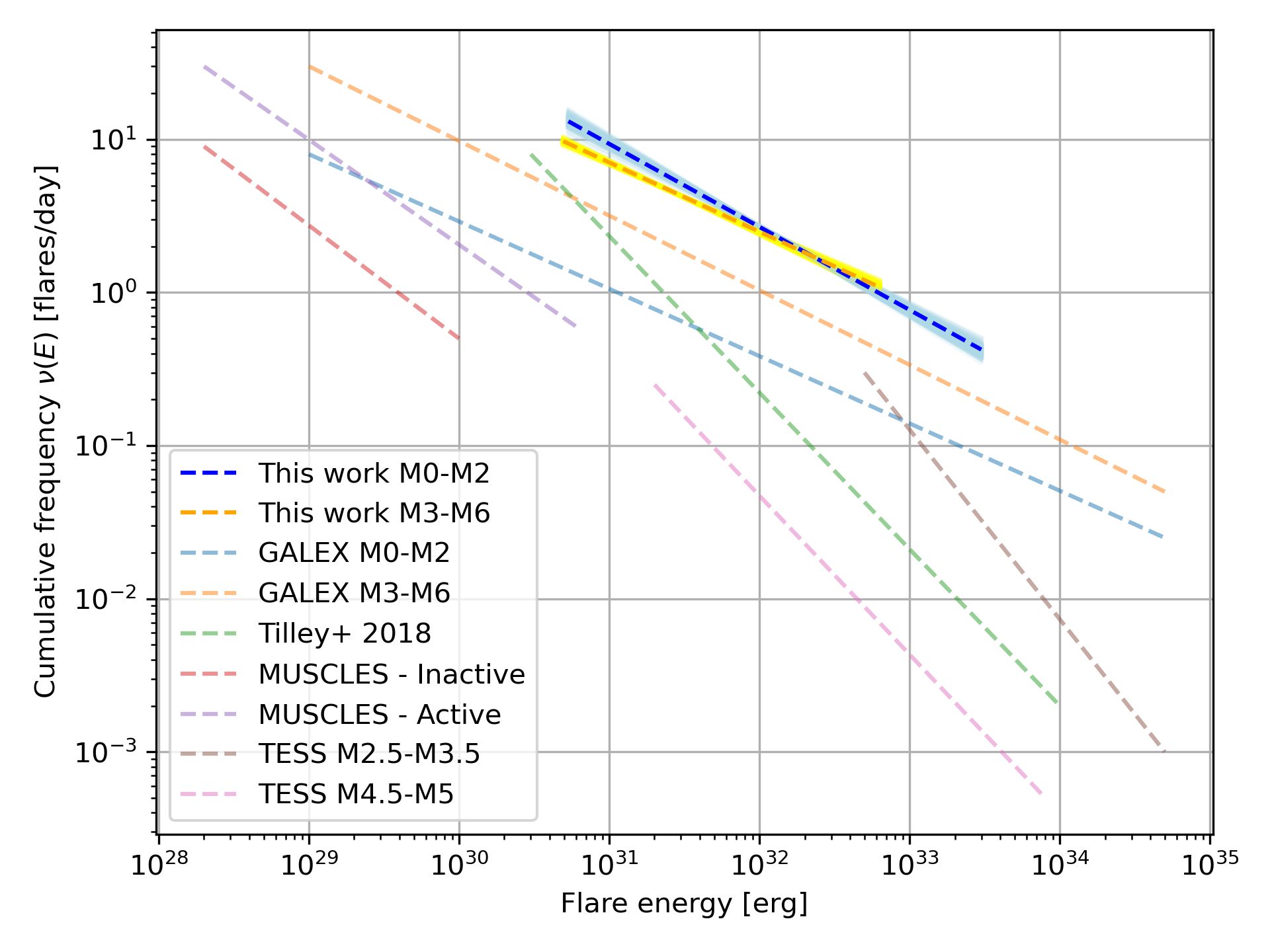}
\caption{Left panel: the FFD measured for the combined sample (this work+literature), with a power law fit. Right panel: the best fit model we obtain compared with previous measurements of the FFD at other wavelengths.}
    \label{fig:ffd_cumulative}
\end{figure*}

In Figure~\ref{fig:ffd_cumulative}, we show the FFD we measured for spectral types M0–M2 and M3–M6. This division, also adopted by \citet{Rekhi2023}, is motivated by the transition to full convection beyond spectral types M3–M4, which may be associated with different flaring behaviors.


Two instrumental selection effects are likely to bias the FFD: (1) the sensitivity of the instrument limits the accessible survey volume, causing low-energy flares to be underrepresented in our sample (Malmquist bias); and (2) the limited exposure time of Chandra restricts the detectable flares to those shorter than the exposure duration, potentially leading to an over-representation of shorter flares in the sample. This second effect combines with the energy--duration correlation discussed in ~\S\ref{sec:energy-duration}. 

To account for these selection biases, we restrict our fits to flares with energies comprised in the interval [$4\times10^{30}, 6\times10^{33}$]. A derivation of this criterion is presented in the Appendix~\ref{selection_function_flare}. An additional selection bias, which we do not attempt to correct in this work, is the under-representation of late type M dwarfs in the SUPERBLINK catalog, unless they are particularly bright or active (due to the cuts $J<10$ and $V-J>2.7$ that were applied when making this catalog).

We estimate the errors on $\nu$ with bootstrapping \citep{2020SciPy-NMeth} and we fit the data using a model similar to that adopted by \citet{Rekhi2023}:
\begin{equation}
\nu(E) = C E^{-\alpha} \,; \qquad C > 0,\ \alpha > 1 ,
\end{equation}
and report the best-fit parameters in Table~\ref{tab:ffd_energy} (1) when using only the new flares detected in this work and (2) when adding to them the sample of flares reported in the literature.

In Figure~\ref{fig:ffd_cumulative} we show the best fit FFD and compare our result to those derived at other wavelengths in previous works. In particular, our results appear to be in good agreement with those derived at the UV by \citet{Rekhi2023}, specifically for M3-6. The average flare occurrence rate we measure is $\sim 10^{-1}\,\rm ks^{-1}$ (corresponding to $\sim 9$ flares per day). We note that the literature sample used to derive the FFD includes $44$ flares 
 originating from a single star (Au Mic), potentially biasing the inferred flare occurrence rates toward the properties of this particular object.

\subsection{The effect of the Flare Frequency on the habitability of potential exoplanets}
To evaluate the impact of the flare frequency on potentially habitable planets, we model an Earth-analog planet ($M_p = 1\,M_{\oplus}$, $R_p = 1\,R_{\oplus}$) orbiting each star at the Earth-equivalent distance, defined as the orbital separation at which a planet receives the same bolometric flux as Earth from the Sun and calculated as

\begin{equation}
a_{\mathrm{EE}} =
\sqrt{\frac{L_\star}{L_\odot}}
\,\mathrm{AU},
\end{equation}
where $L_\star$ is the stellar bolometric luminosity.
Stellar luminosities were assigned using the calibration of \cite{mamajek2013}. 

The incident flux received from the planet is calculated as:
\begin{equation}
F_{\mathrm{flare,planet}} =
F_{\mathrm{flare}}
\left(
\frac{d}{a_{\mathrm{EE}}}
\right)^2,
\end{equation}
where $d$ is the distance at which the flare is observed.

We checked that the Earth-equivalent orbital distances lie within the habitable zone boundaries defined by the Runaway Greenhouse and Maximum Greenhouse limits of \cite{kopparapu2013, kopparapu2014}. Atmospheric escape is calculated using the hydrodynamic calculation of \cite{amaral2025}. At low incident extreme ultraviolet and X-ray (XUV) fluxes, also refered to at the ``Energy Limited'' (EL) regime, the mass-loss rate $\dot{M}_{\mathrm{EL}}$ is given by \citep{erkaev2007, luger2015}:
\begin{equation}
\dot{M}_{\mathrm{EL}} =
\frac{
\eta \pi F_{\mathrm{XUV}} R_p R_{\mathrm{XUV}}^2
}{
G M_p K_{\mathrm{tide}}
}
\end{equation}
where $\eta$ is the hydrogen escape efficiency due to the XUV, $R_{\mathrm{XUV}}$ is the radius of influence (measured from the center of the planet, which indicates how much the XUV can penetrate into the atmosphere of the planet), and $K_{\mathrm{tide}}$ is the parameter that accounts for tidal effects in the atmosphere. $F_{XUV}$ is the XUV flux incoming to the planet, $M_{p}$ is the planetary mass, $R_{p}$ is the planetary radius and G is the gravitational constant. 
At high fluxes, atmospheric escape transitions to the ``Radiation Recombination'' (RR) regime, where the mass-loss rate $\dot{M}_{\mathrm{RR}}$ is  by  \citep{murrayclay2009, luger2015}:
\begin{equation}
\dot{M}_{\mathrm{RR}} =
2.248 \times 10^{6}
\sqrt{
F_{\mathrm{XUV}} R_p^3
}
\end{equation}

The transition between the two regimes occurs at the critical flux given in~\S\,6 of \cite{amaral2025}.
Following this model, we adopt 
a hydrogen escape efficiency typical of hydrogen-rich atmospheres \citep{luger2015}. 
Following the methodology of \cite{amaral2025} for terrestrial-sized planets around active M dwarfs, we assume $R_{\rm XUV} = R_{\rm p}$ 
(which provides a lower bound on the mass loss rate). 

For the tidal correction factors, we use $K_{\rm tide} = 0.95$, in agreement with the range of the values provided \cite{luger2015} for terrestrial planets in the habitable zones of M dwarfs ($0.9$ to $0.99$).

We note that our observations only include X-ray wavelenghts, and since  $F_{X}<F_{XUV}$, the values we find for $\dot{M}_{\mathrm{EL}}$ and $\dot{M}_{\mathrm{RR}}$ are lower limits.




To estimate the cumulative atmospheric erosion driven by stellar flaring activity, we compute the time required for an Earth-like atmosphere
 ($M_{\mathrm{atm}} \approx 5 \times 10^{21}\,\mathrm{g}$; \citealt{trenberth2005})
to be removed through hydrodynamic escape.
The cumulative flare-driven atmospheric mass loss is estimated by integrating the FFD derived in \S~\ref{sec:rates} over the observed flare energy range. 
Flare durations are scaled with flare energy using the empirical relations computed in \S~\ref{sec:energy-duration}, consistent with X-ray flare scaling relations \citep{veronig2002, christe2008}. We find that the upper limit on the atmospheric erosion time spans $0.5$ --$30$\,Myr
, depending on the spectral type of the host star, with this timesclae growing linearly with $M_{atm}$. 

We note that recent works (e.g. \citealt{caldiroli2025}) suggest that these numbers may have to be considered cautiously and that atmospheric survival remains highly sensitive to the interplay between the specific flaring environment and orbital distance \citep{amaral2025}. More extensive long-term X-ray monitoring of M dwarfs, to characterize quiescent emission levels and flare occurrence rates across stellar populations spanning a range of ages, metallicities and rotational velocities are necessary to refine these numbers.



\section{Summary}\label{sec:conclusion}

In order to exhibit a picture of the parameters space of X-ray flares from M dwarfs, we did the following: 
\begin{enumerate}
\item We compiled the sample of known events from the literature. We found $122$ flares from $15$ stars (three of which are binaries) which were observed for more than $1350\,\rm ks$ (this number is a lower limit based on the cases where exposure times were provided). Among these flares, $46$ ($38\%$) originated from the same star, Au\,Mic. 
\item We cross-matched the catalog by \cite{Magaudda2022} (the eROSITA observations of the  SUPERBLINK targets), with twenty-five years of Chandra archival observations. This cross-match revealed $7$ additional flaring M dwarfs, observed for a cumulative exposure time of $620\,\rm ks$.
\end{enumerate}

We analyzed the properties of the flares of this combined sample. Our main findings are listed bellow.
\begin{enumerate}
\item The X-ray flares in the combined sample span four orders of magnitude in energies: from $10^{29}\,\rm erg$ to $10^{33}\,\rm erg$. They exhibit a strong correlation between the flare duration and the flare  strength, a trend that is predicted by theory and was observed in the Sun \citep[e.g.,][]{veronig2002, christe2008}.
\item During flares, the fluxes we observe increase by factors of $2.1$ to $22.1$ and the temperature we observe increase by factors of $1.1$ to $5.5$, consistent with theoretical expectations. 
\item The flares have an average duration of $3.7$ ks, with  $1\sigma=2.20\,\rm ks$. The current data favors asymmetric temporal profiles with longer decay phases than rise phases.

\item We derived constraints on the flare frequency distributions of M0–M6 stars up to energies of $10^{33}\,\rm erg$. The average flare occurrence rate we measure is $\sim 10^{-1}\,\rm ks^{-1}$. We also note that $50\%$ of the M dwarfs observed to flare in X-rays spend less than $\sim6\%$ of the total observed time in a flaring state. The inferred rates are in good agreement with the most recent constraints derived in the UV, despite a sample smaller by an order of magnitude. We derived these constraints over an energy interval designed to minimize selection biases, but future large-scale, systematic surveys with better quantified completeness criteria, e.g. using Gaia, are expected to confirm and refine our results.

\item Using several of these results and recent simulations by \cite{amaral2025} of flare-driven atmospheric escape, we find that habitable Earth-like planets with Earth-mass atmospheres, orbiting these M dwarfs, would completely lose their atmospheres within $0.5$--$30\,\rm Myr$, depending on the M dwarf spectral type. This range represents an upper limit.

\end{enumerate}

\begin{acknowledgments}

We thank Tsevi Mazeh, Na'ama Hallakoun and Sivan Ginzburg for their precious comments and Qichun Liu for sharing insights. 

This research was supported by the Israeli Science Foundation (grant Nos. 2068/22 and 2751/22)

This research has made use of data obtained from the Chandra Data Archive and the Chandra Source Catalog, both provided by the Chandra X-ray Center (CXC).


This research was supported by Deutsche Forschungsgemeinschaft  (DFG, German Research Foundation) under Germany’s Excellence Strategy - EXC 2121 "Quantum Universe" – 390833306. Co-funded by the European Union (ERC, CompactBINARIES, 101078773). Views and opinions expressed are however those of the author(s) only and do not necessarily reflect those of the European Union or the European Research Council. Neither the European Union nor the granting authority can be held responsible for them.

E.O.O. is grateful for support by grants from the Willner Family Leadership Institute, Madame Olga Klein-Astrachan, André Deloro Institute, Schwartz/Reisman Collaborative Science Program, Paul and Tina Gardner, The Norman E Alexander Family Foundation ULTRASAT Data Center Fund, Jonathan Beare, Israel Science Foundation, Minerva, BSF, BSF-transformative, and the Weizmann-UK.
\end{acknowledgments}
\begin{figure*}
    \centering
    \includegraphics[width=8.5cm]{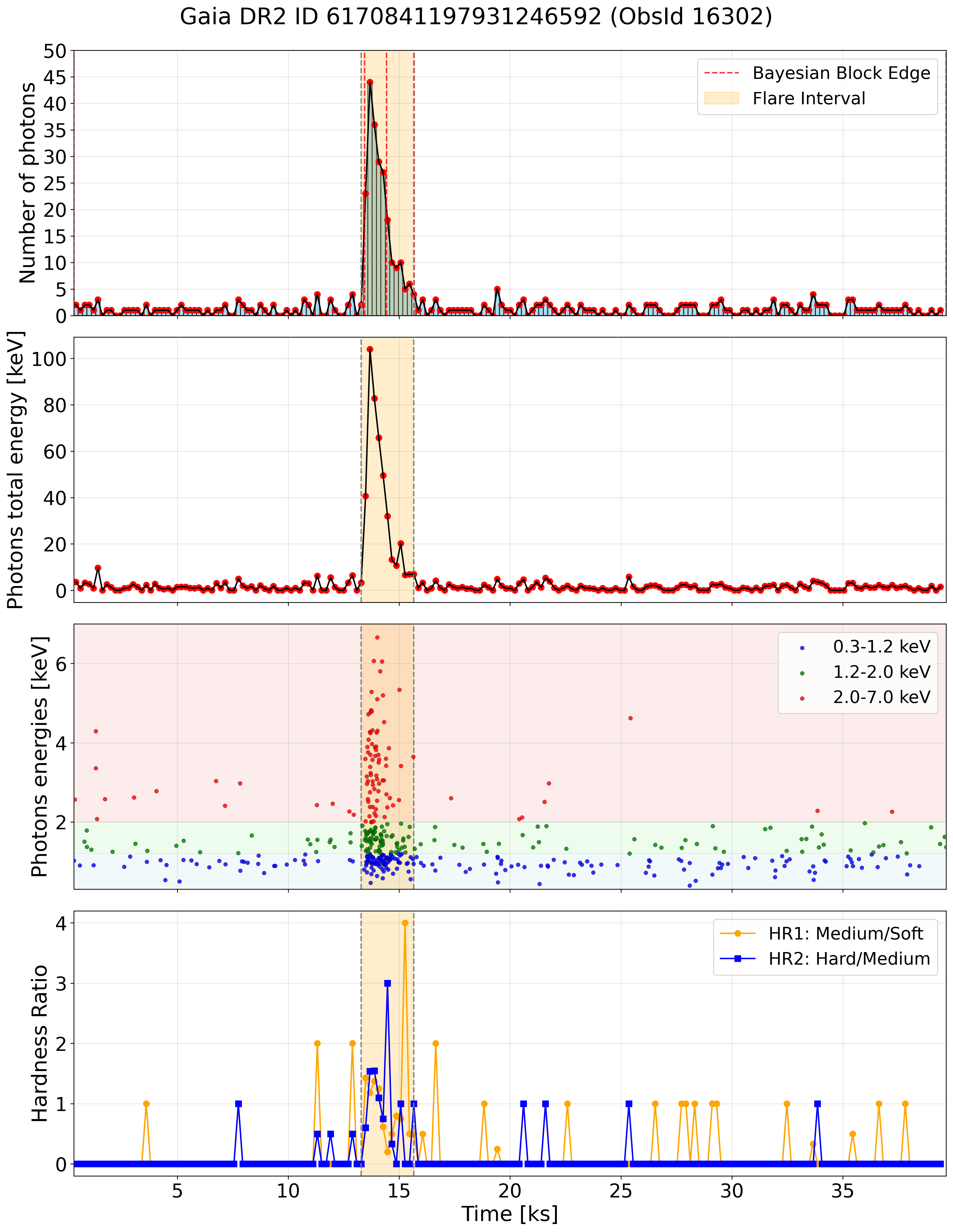}
    \includegraphics[width=8.5cm]{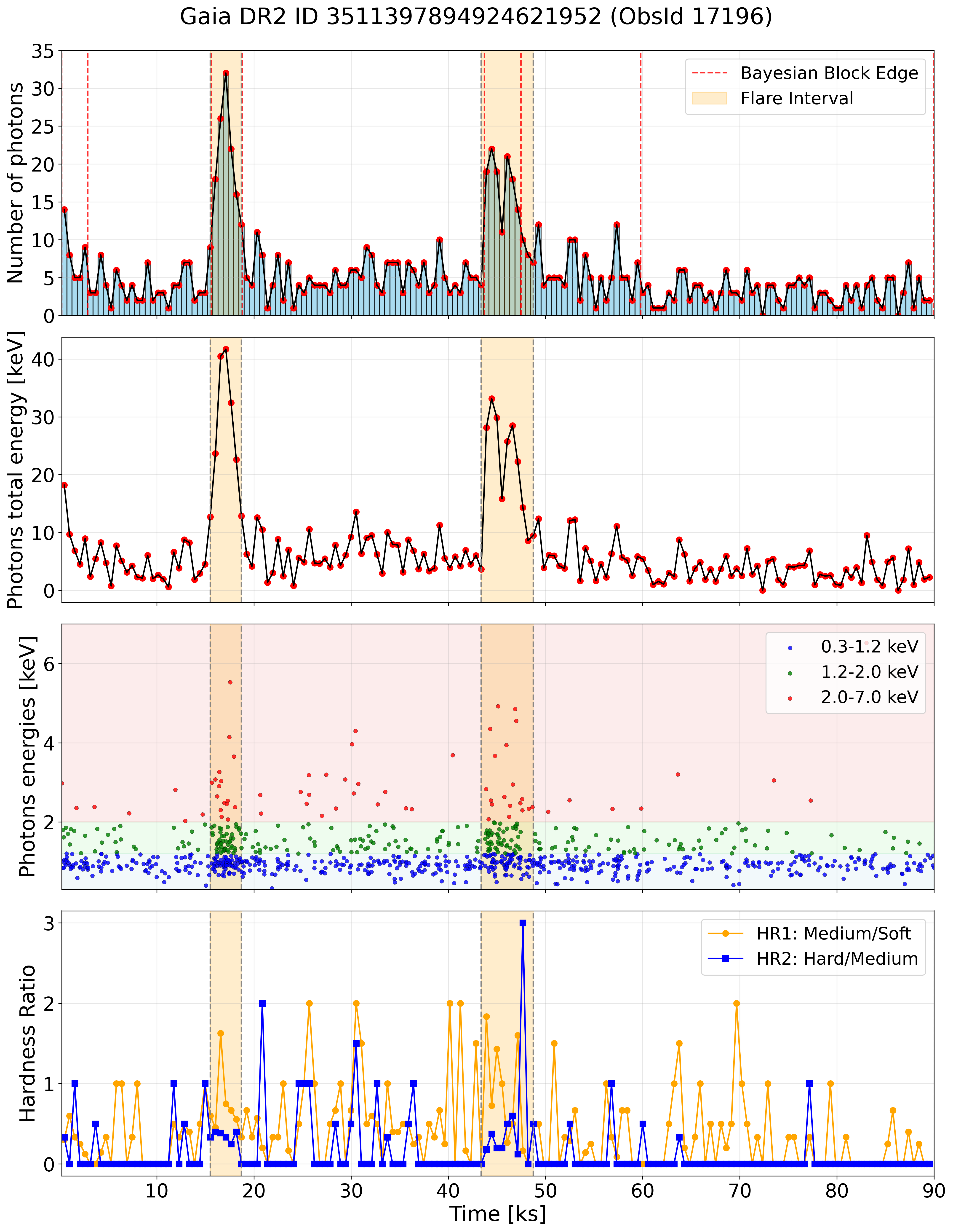}
    \includegraphics[width=8.5cm]{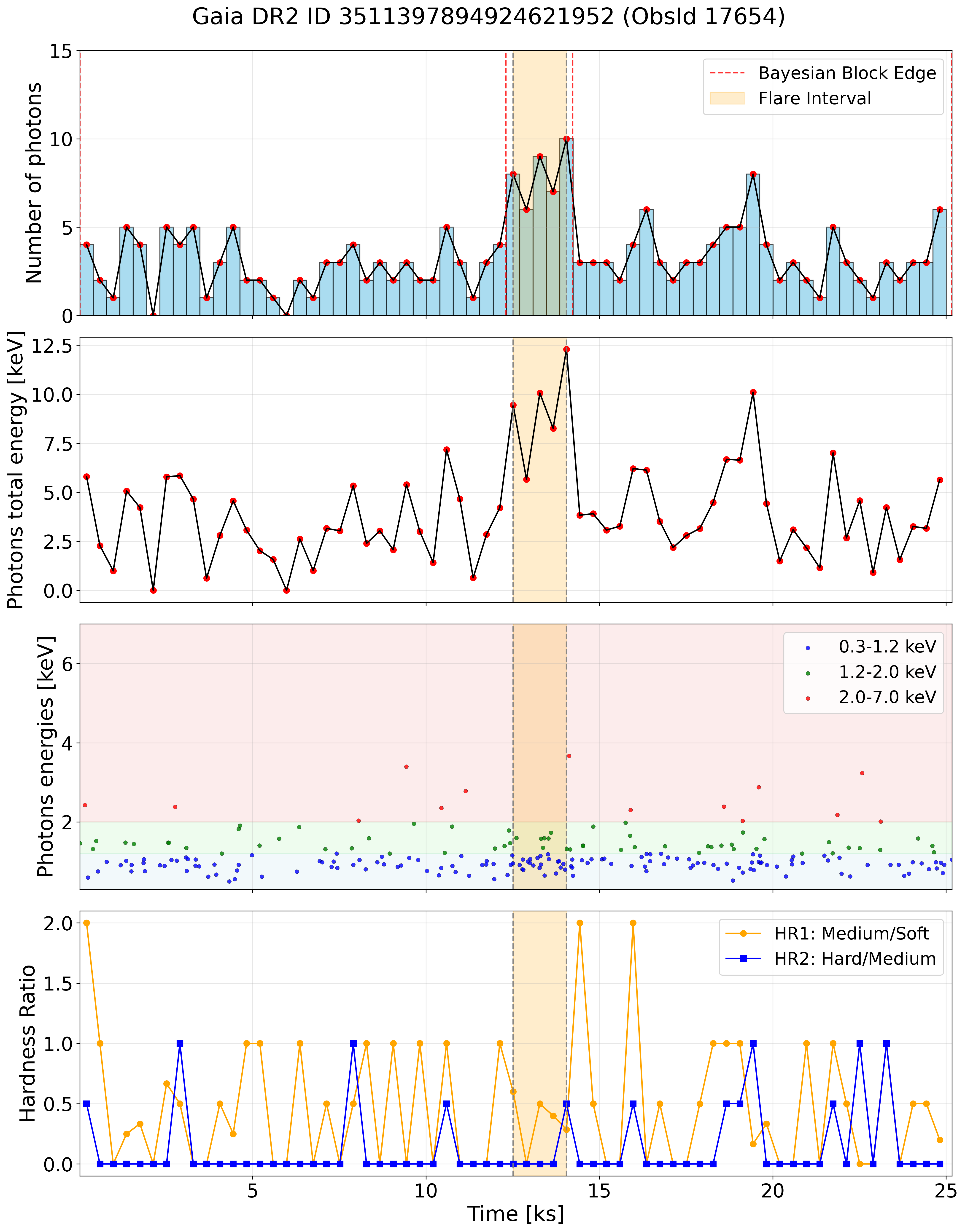}
    \includegraphics[width=8.5cm]{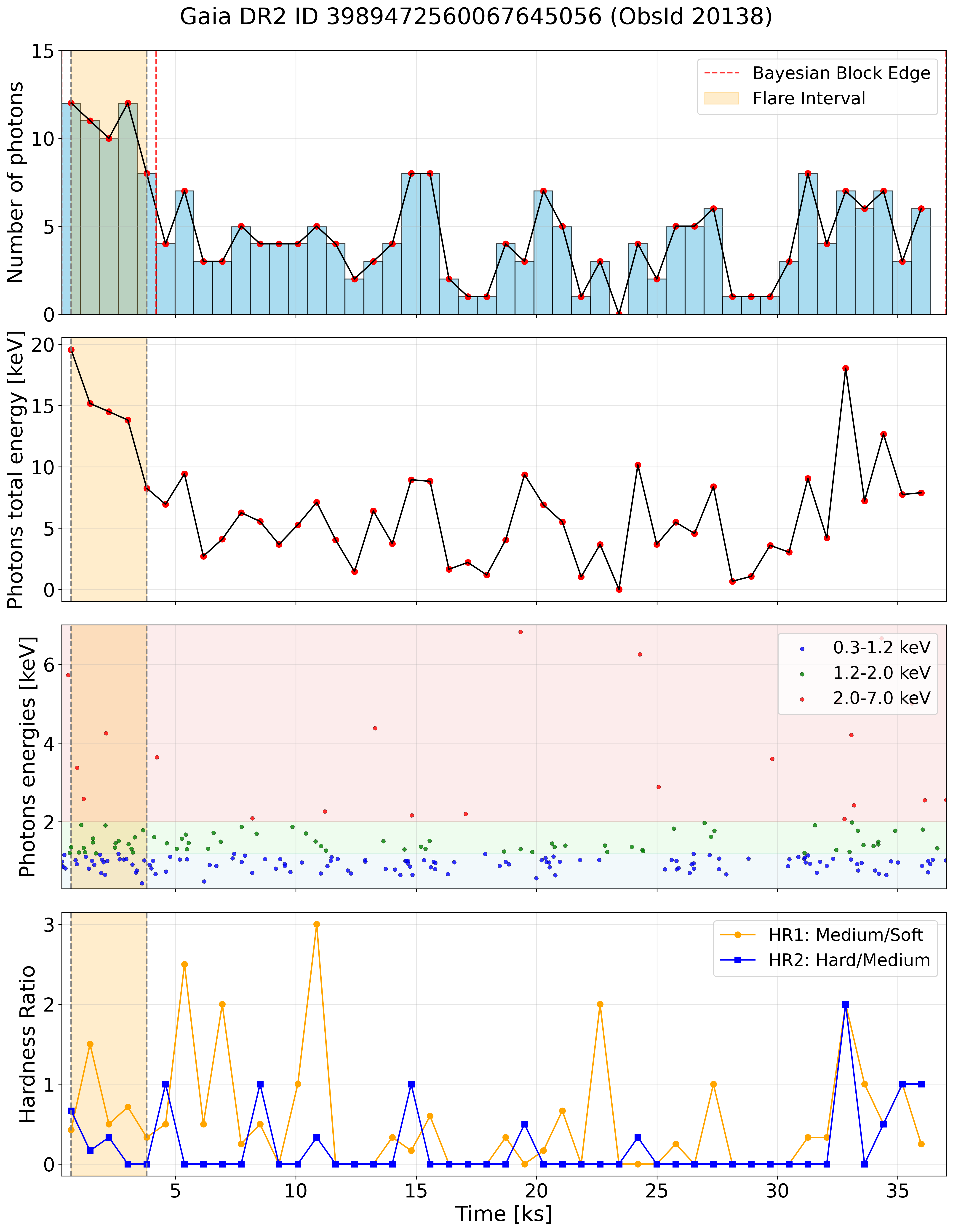}
\end{figure*}

\begin{figure*}
    \centering
    \includegraphics[width=8.5cm]{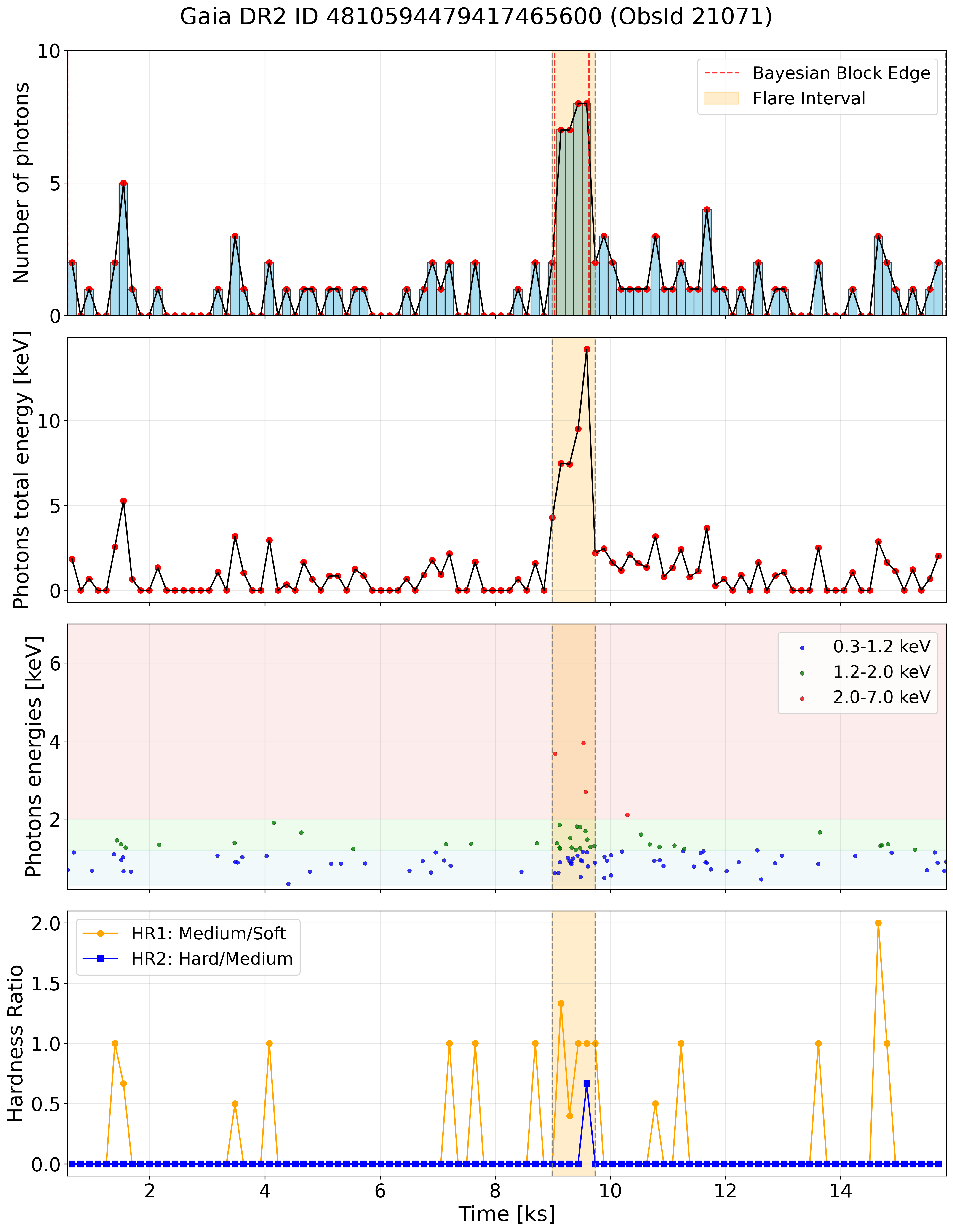}
    \includegraphics[width=8.5cm]{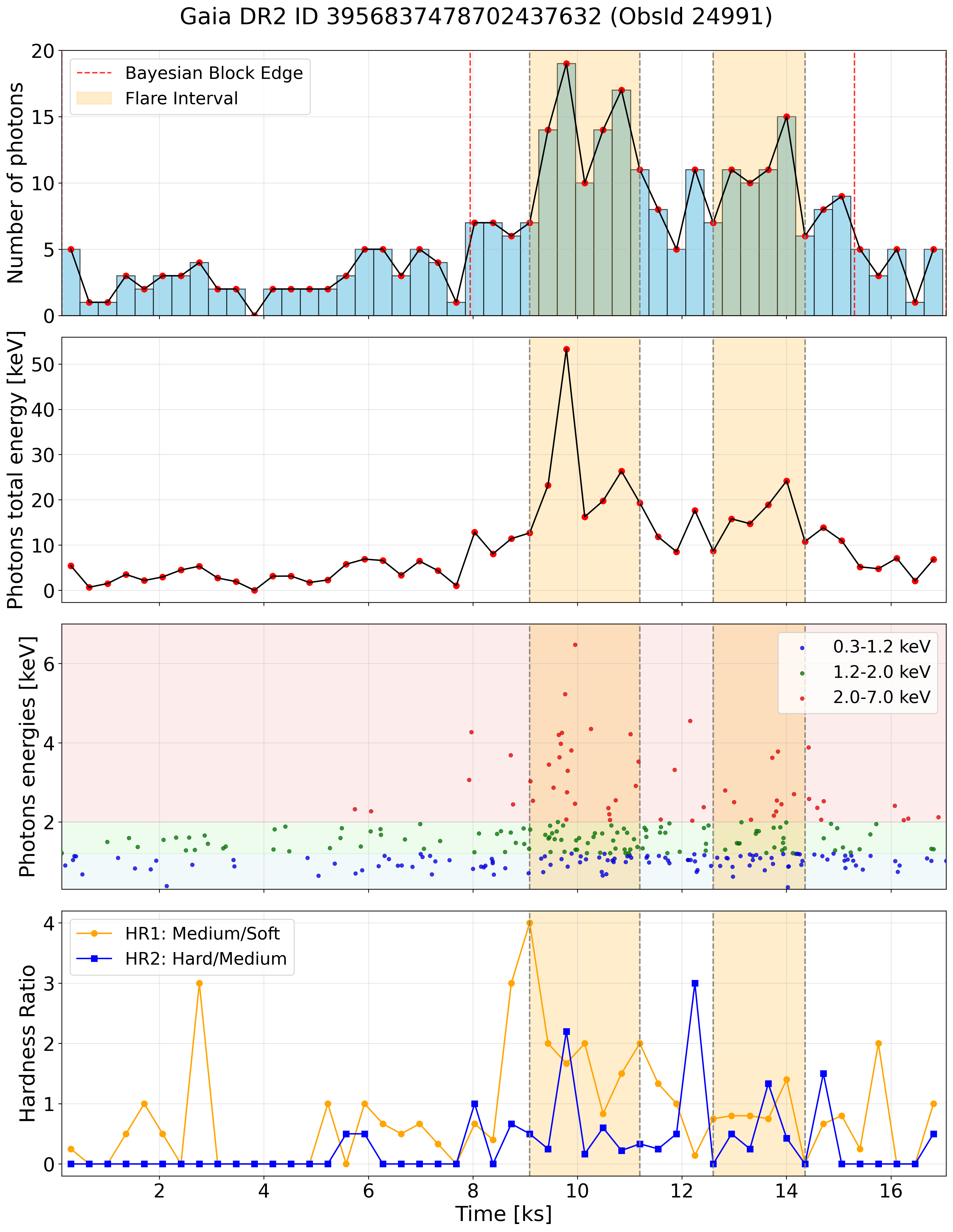}
    \includegraphics[width=8.5cm]{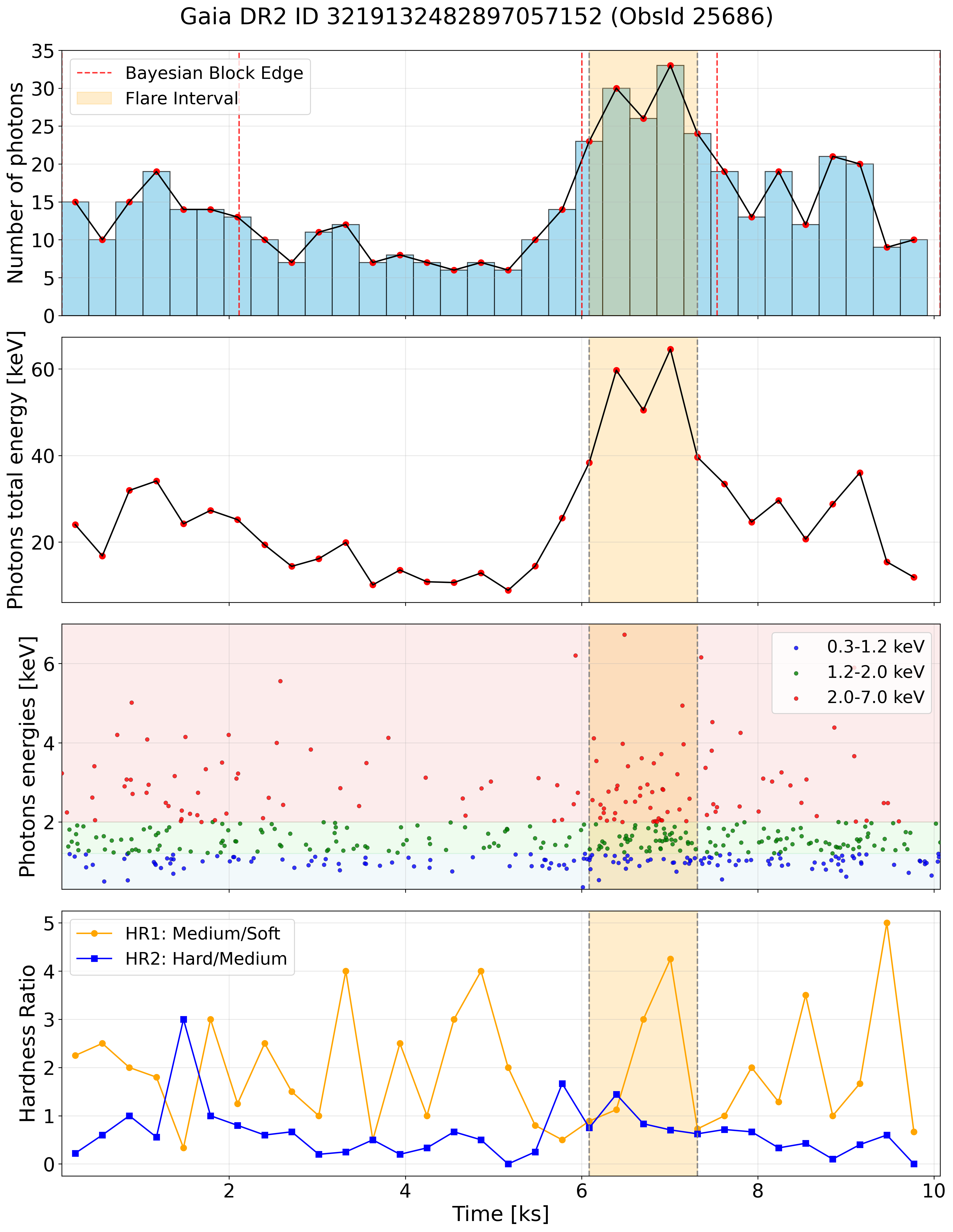}
    \includegraphics[width=8.5cm]{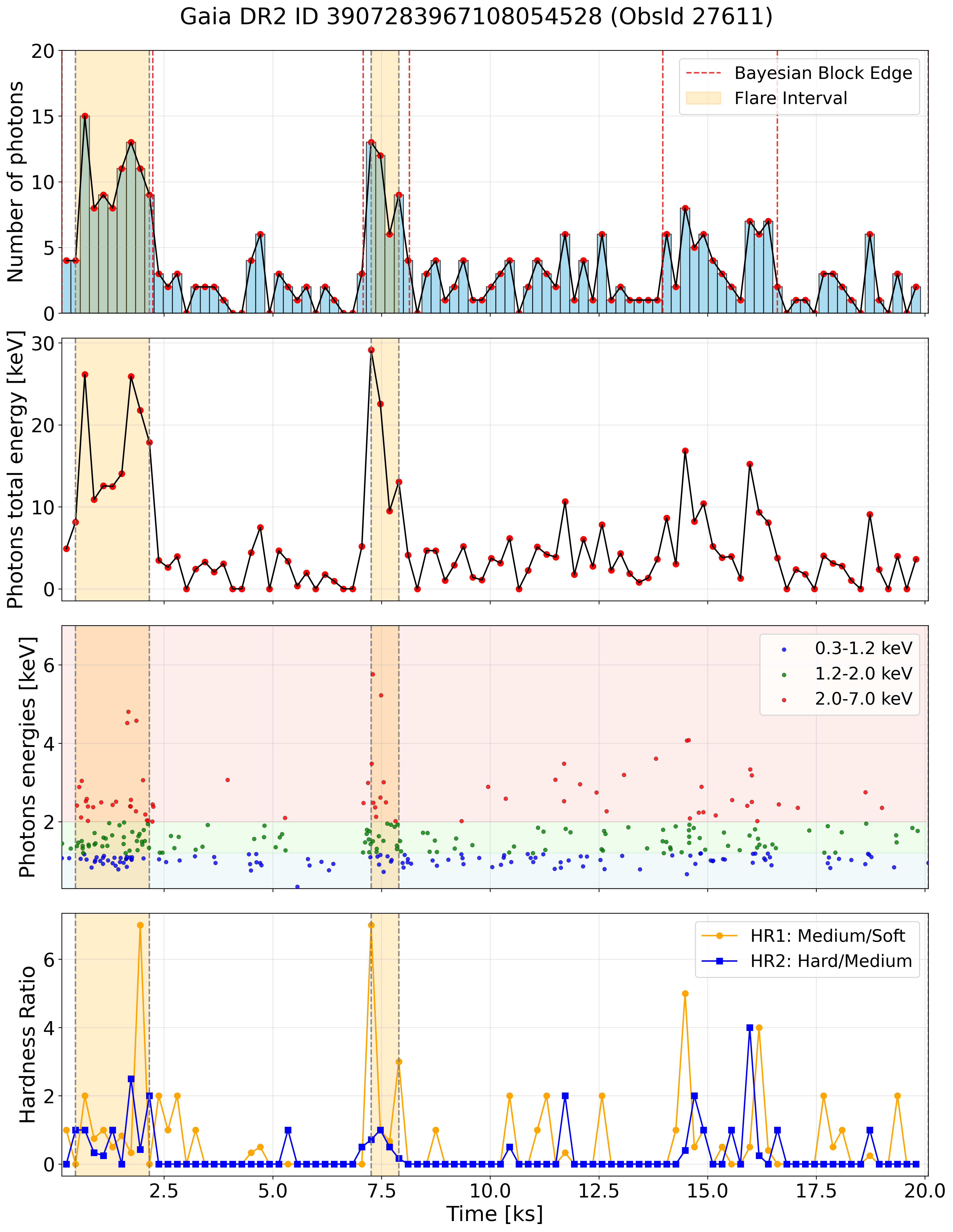}
\end{figure*}
\begin{figure*}
    \includegraphics[width=8.5cm]{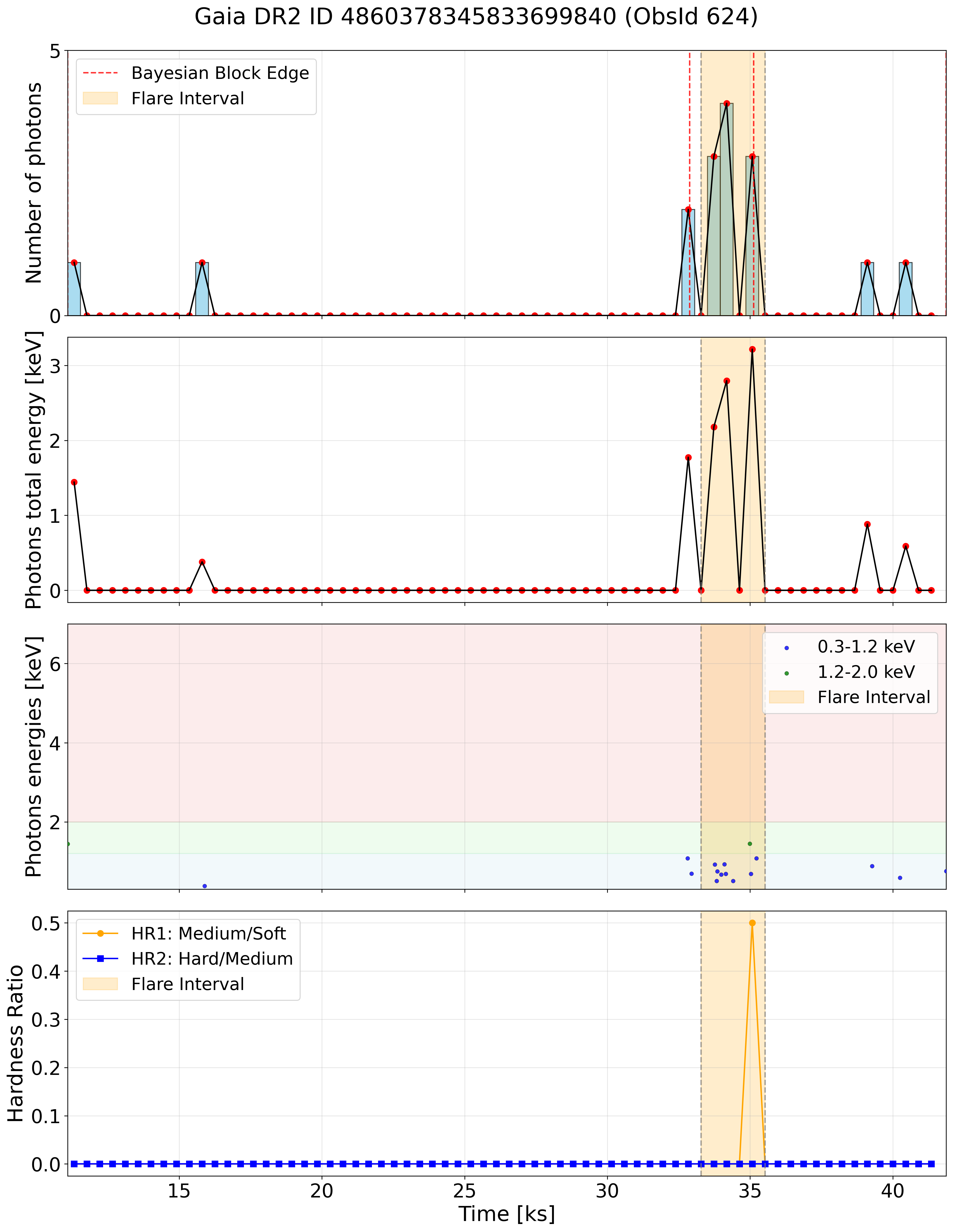}
    \caption{
 {\bf Light curves of the flares}. All panels share the same x-axis, showing photon arrival times measured from the beginning of each observation. From top to bottom, the panels display: (1) the number of detected photons, (2) the total photon energy per time bin, (3) individual photon energies (keV), and (4) the hardness ratio as defined in \S\ref{sec:hardness_ratio}. Regions shaded in orange indicate the time intervals during which a flare was identified by our algorithm.
}
    \label{fig:combined}
\end{figure*}

\begin{figure*}
    \centering
    \includegraphics[width=7cm]{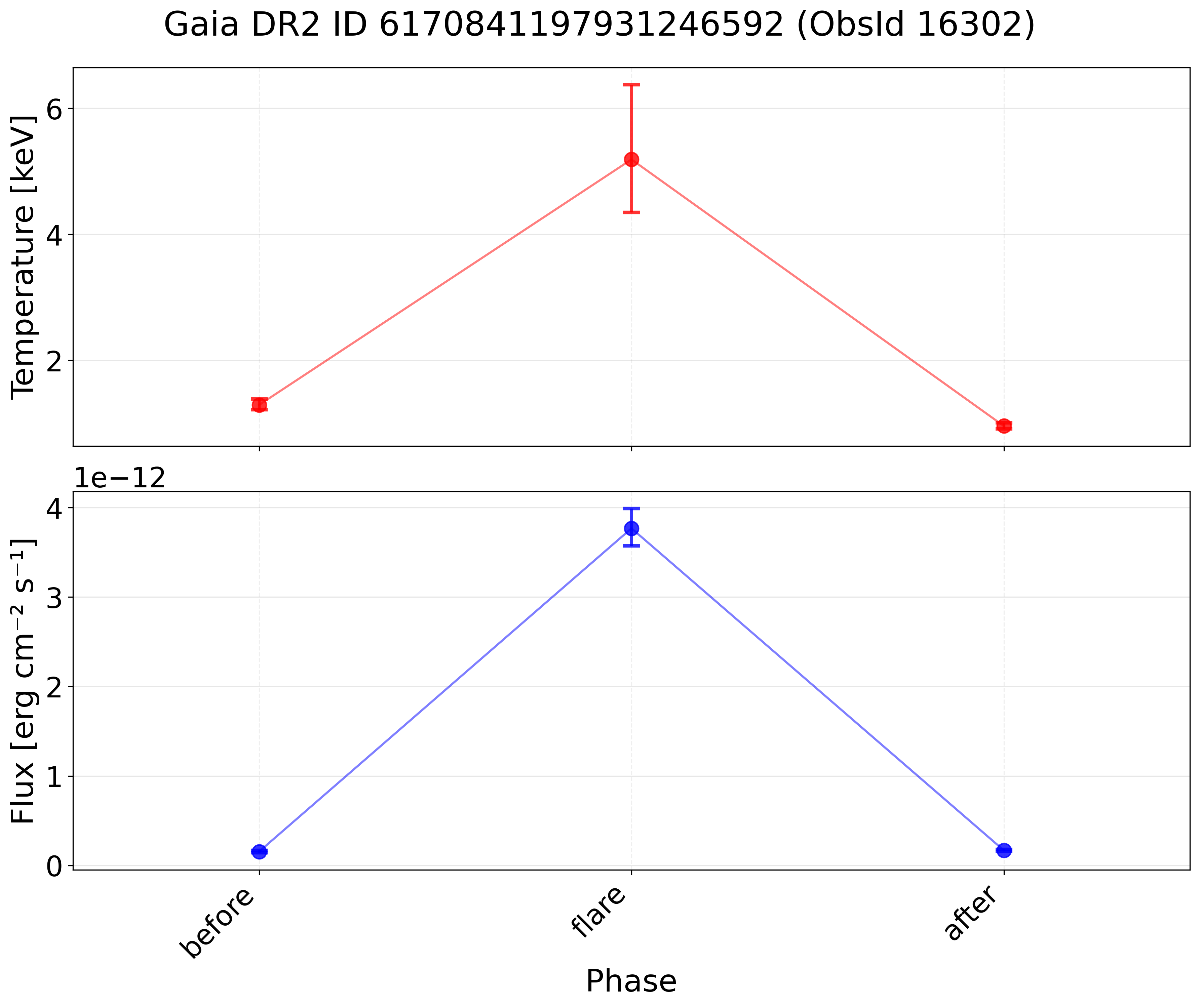}
    \includegraphics[width=7cm]{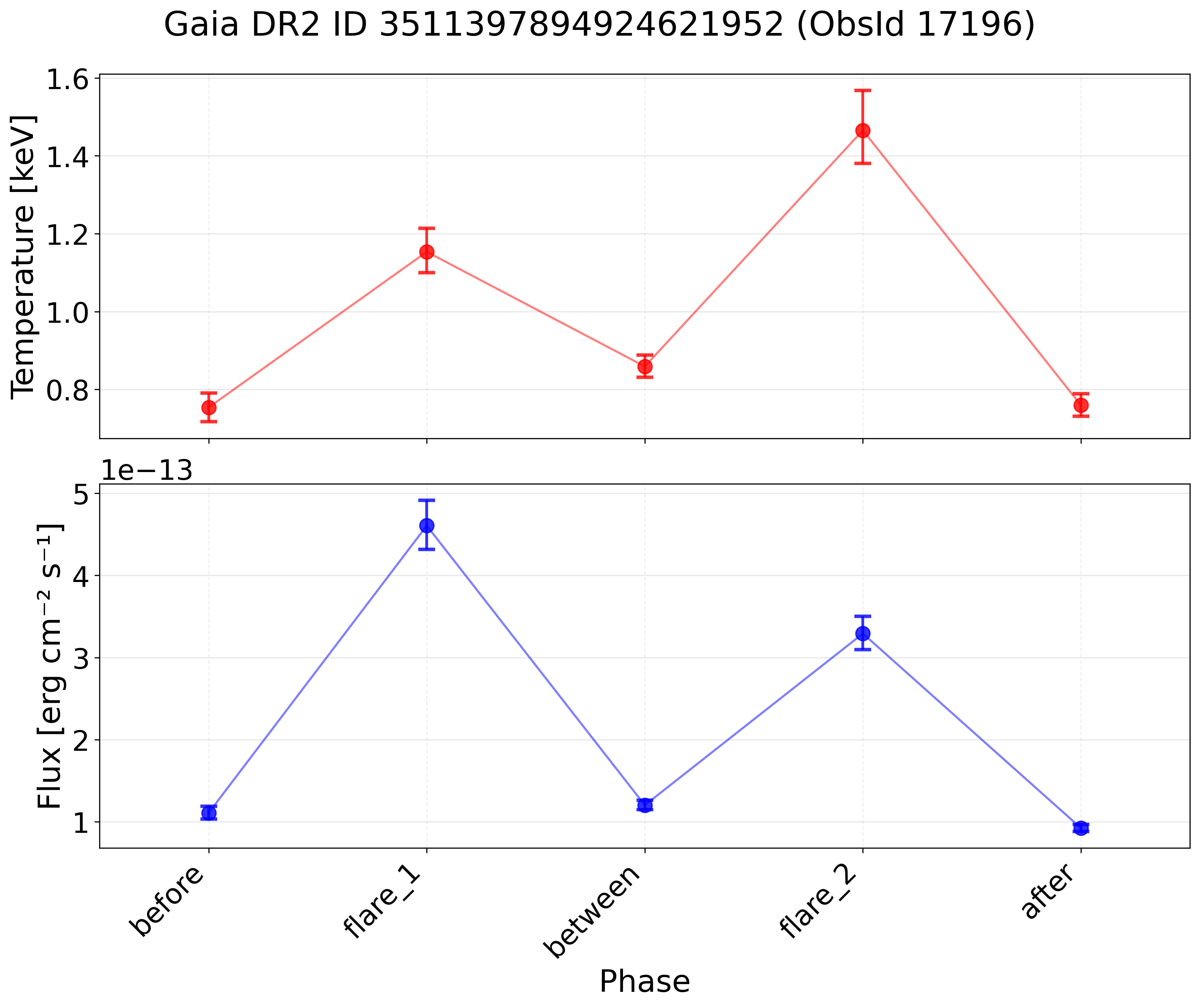}
    \includegraphics[width=7cm]{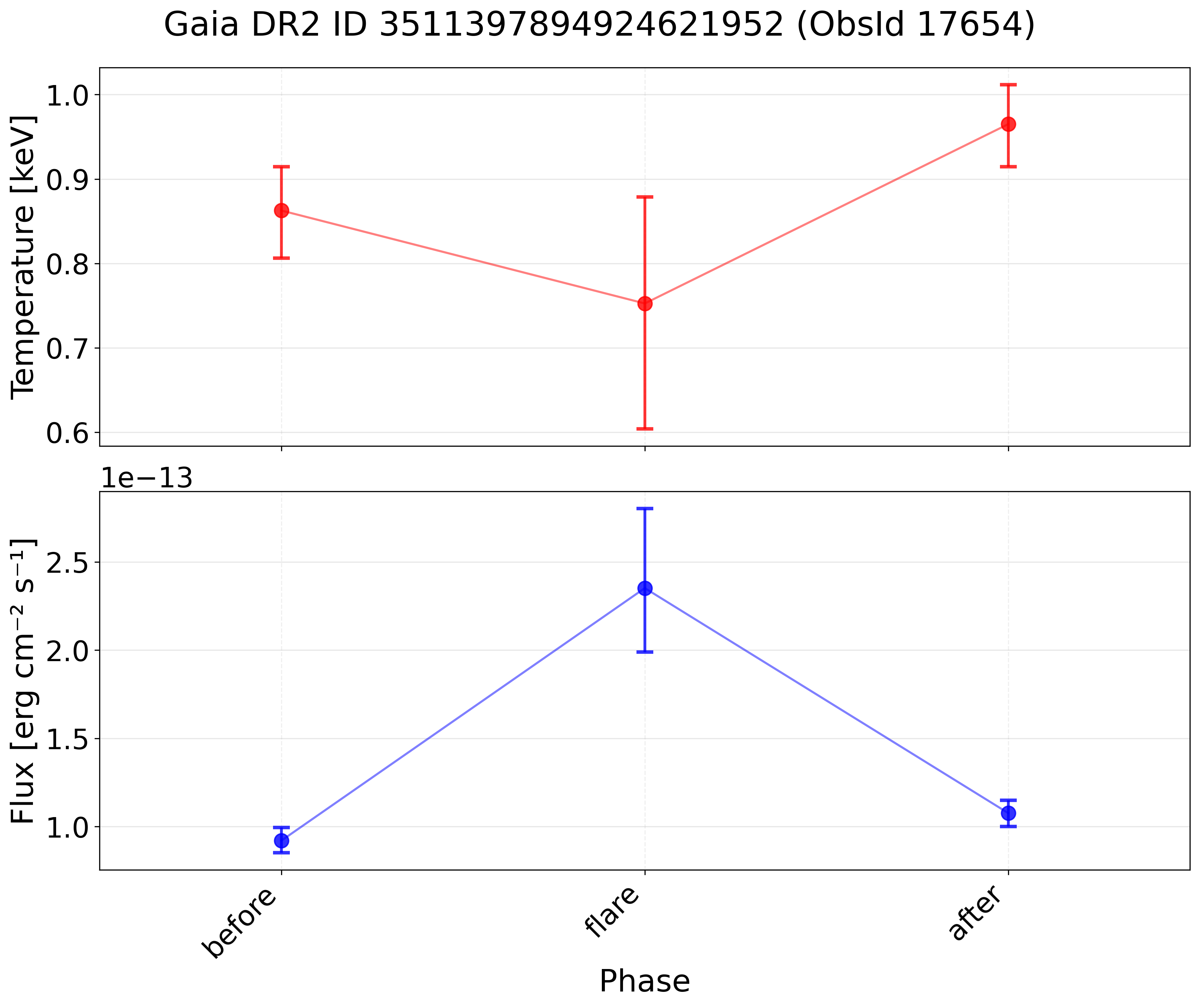}
    \includegraphics[width=7cm]{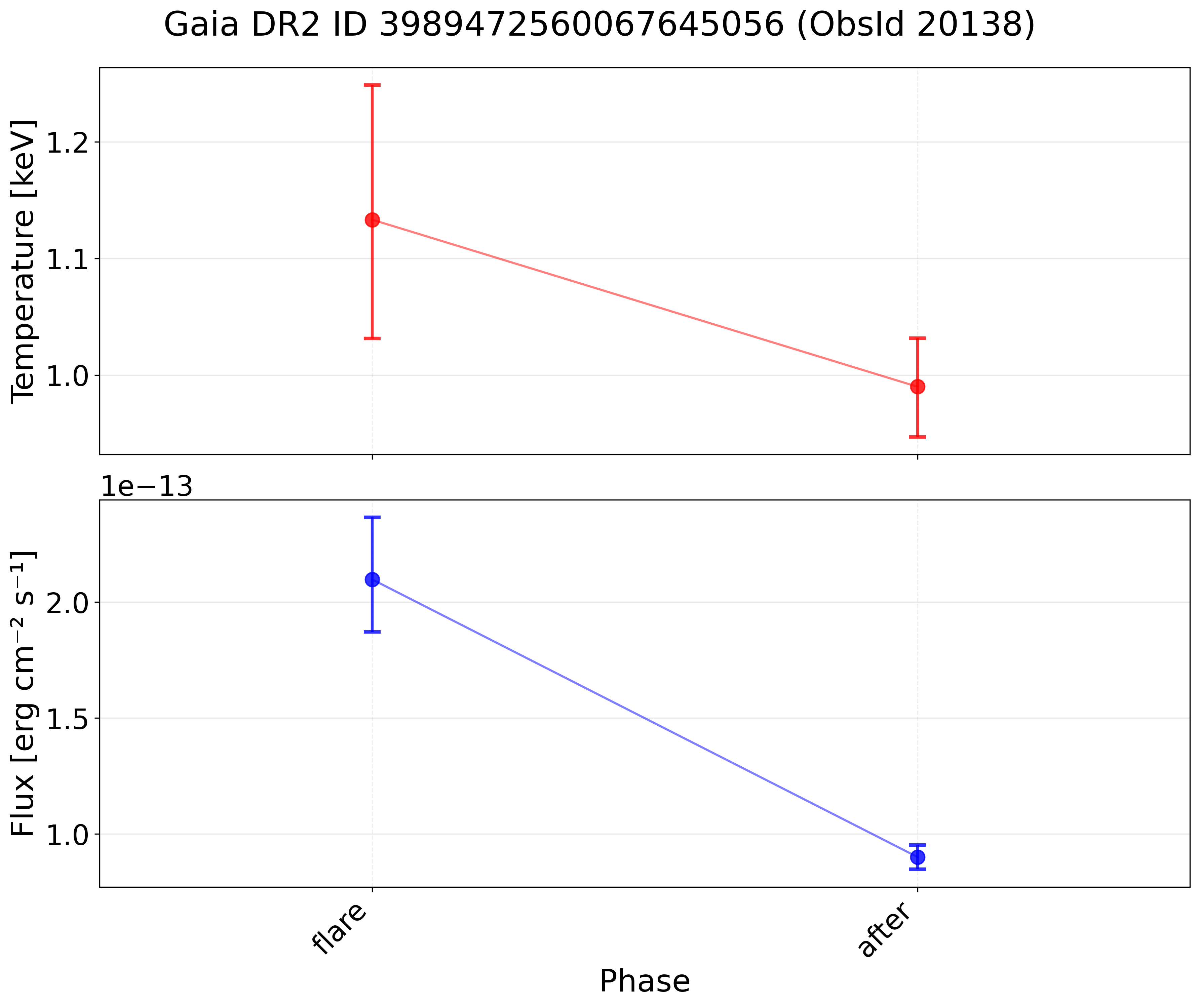}
    \includegraphics[width=7cm]{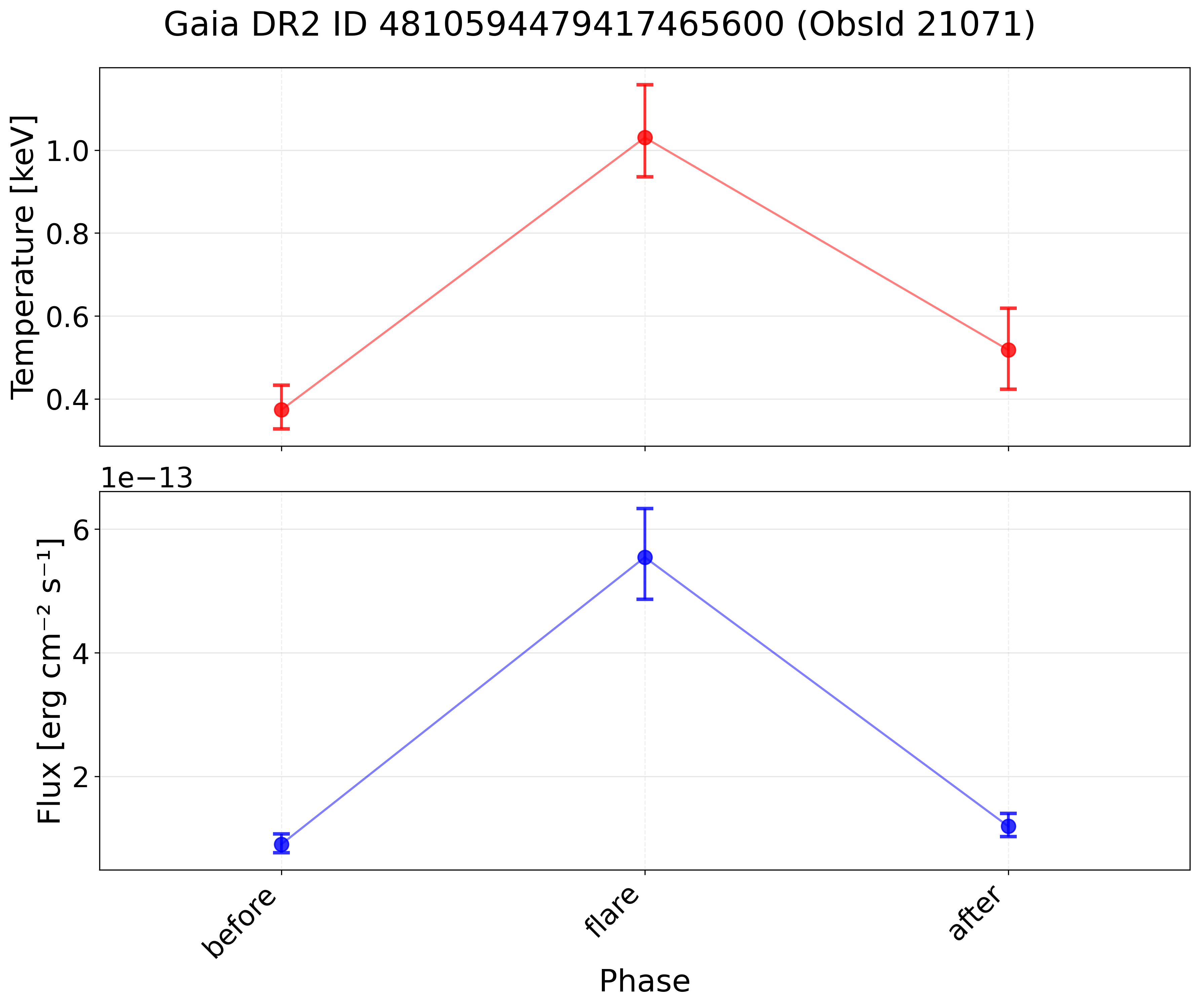}
    \includegraphics[width=7cm]{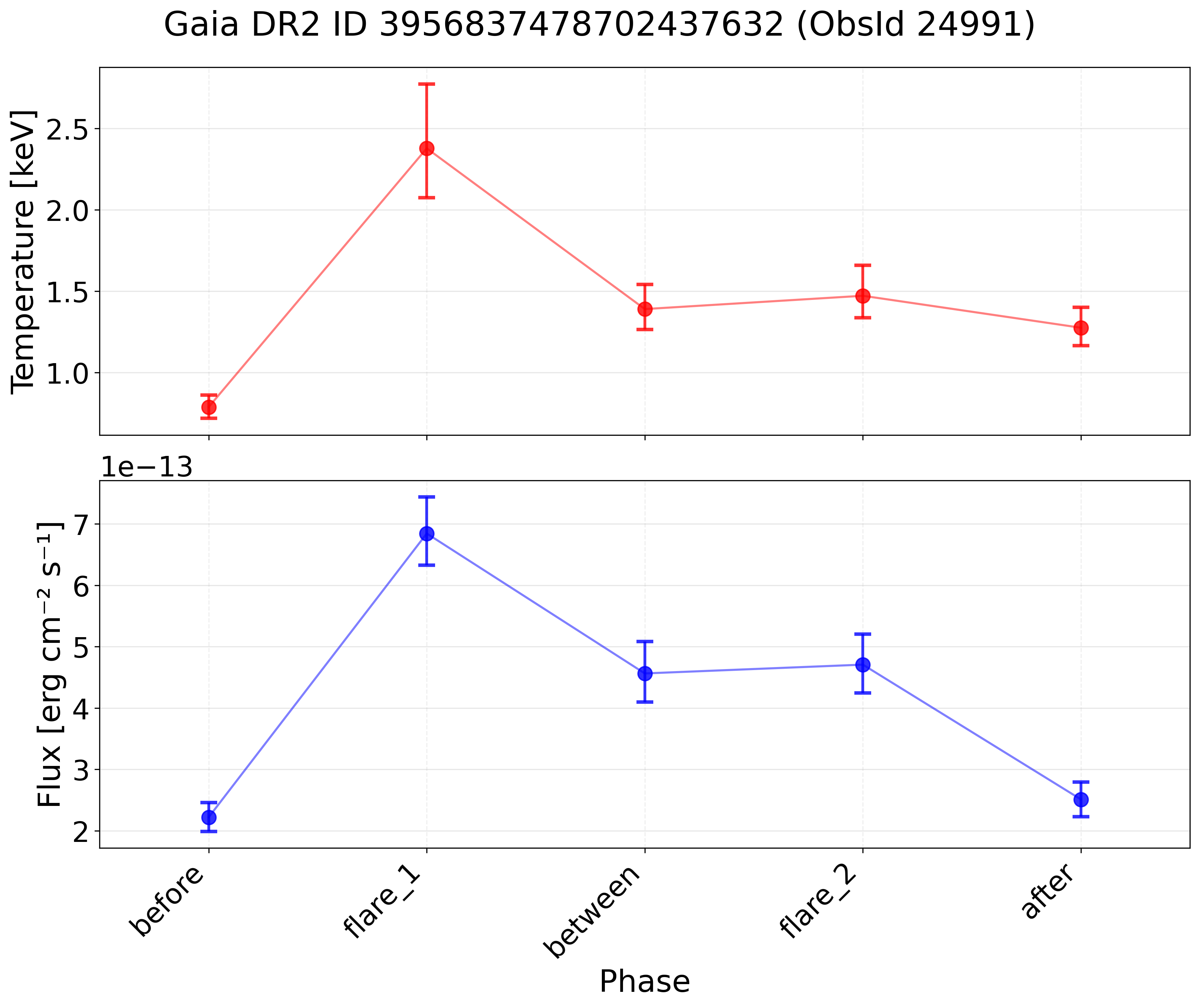}
    \includegraphics[width=7cm]{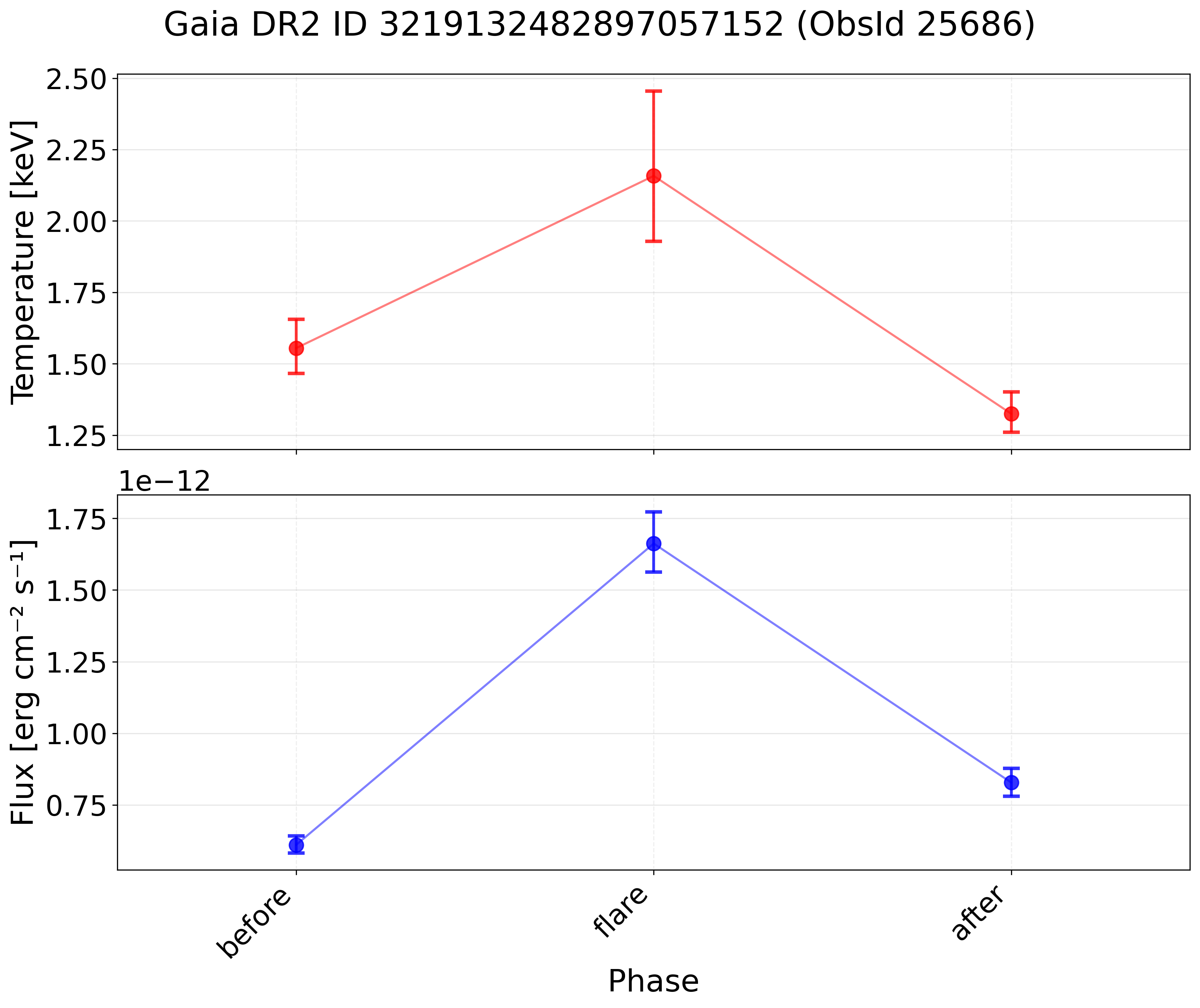}
    \includegraphics[width=7cm]{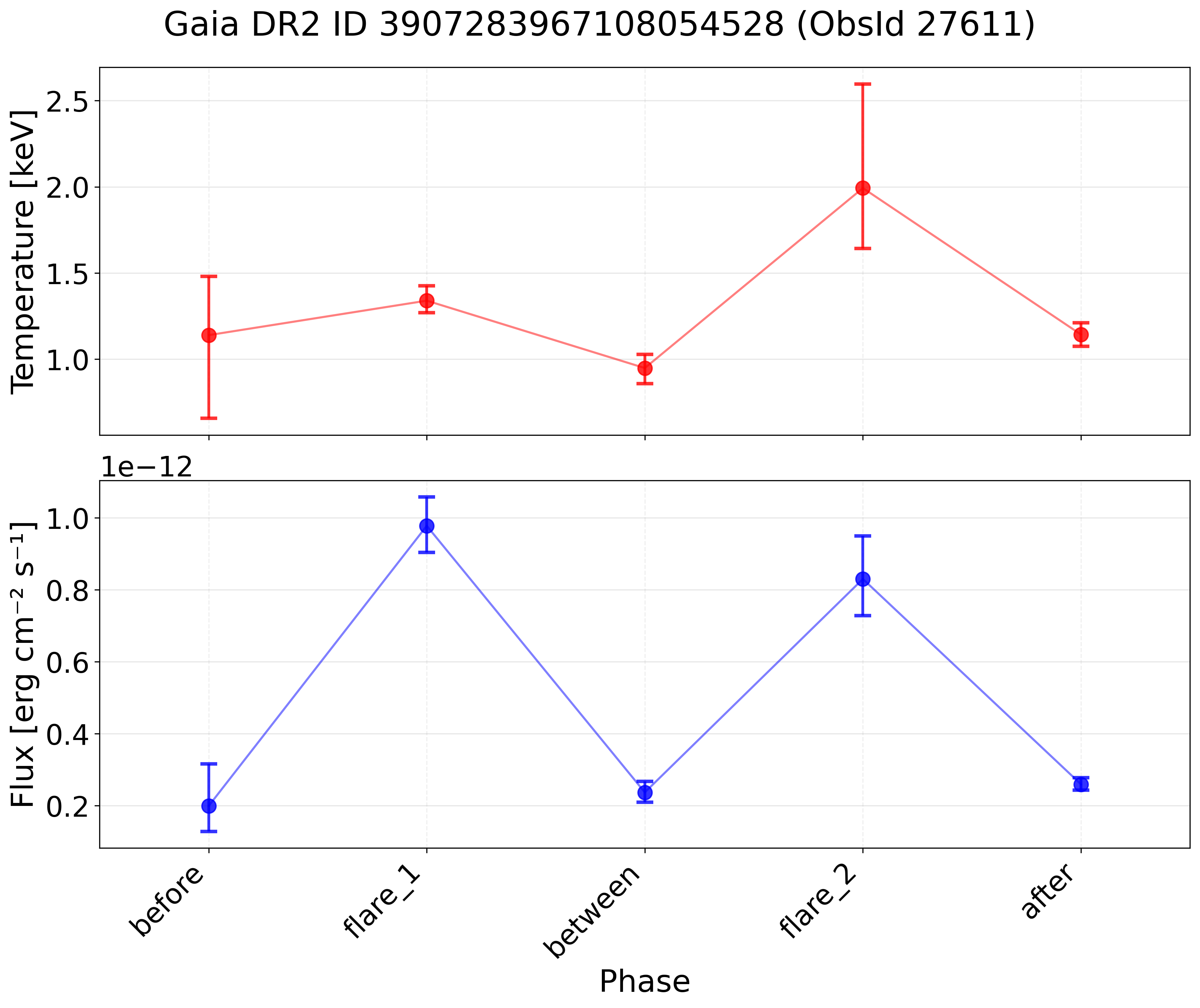}
    \caption{{\bf Evolution of the temperature and flux of the flares}. For each observation, the upper panel shows the plasma temperature, and the lower panel shows the X-ray flux, before, during and after the flare. Both quantities were derived with the {\tt XSPEC} spectral fitting code. The different phases (before-during-after) were defined using the Bayesian block algorithm described in~\S\ref{sec:binning}.
}
    \label{fig:temp_flux}
\end{figure*}

\appendix

\section{Estimation of Selection bias effects in the FFD measurements}\label{selection_function_flare}

The main selection effects that affect the FFD are: (1) the sensitivity of the instrument limits the accessible survey volume, causing low-energy flares to be underrepresented in our sample (Malmquist bias); and (2) the limited exposure time of Chandra restricts the detectable flares to those shorter than the exposure duration, potentially leading to an over-representation of shorter flares in the sample. This second effect combines with the energy--duration correlation discussed in ~\S\ref{sec:energy-duration}.

Here, we explain how we derive an energy interval, $[E_{\rm min},E_{\rm max}]$, over which the flare sample is expected to be relatively complete. We note that we did not attempt to correct for faint late-type M dwarfs being under-represented in the SUPERBLINK catalog.

For the instrumental sensitivity limit, we used  
$F_{\rm lim}=10^{-14}~{\rm erg~s^{-1}~cm^{-2}}$
\citep{evans2024}, consistent with deep X-ray field analyses \citep{Magaudda2022}.

Assuming the flare duration–energy relation
\begin{equation}
T=\tau_0\left(\frac{E}{E_0}\right)^{\beta},
\end{equation}

the observed flare flux writes
\begin{equation}
F=\frac{E^{1-\beta}E_0^\beta}{4\pi d^2\tau_0}
\end{equation}

\begin{figure*}
\centering
\includegraphics[width=8cm]{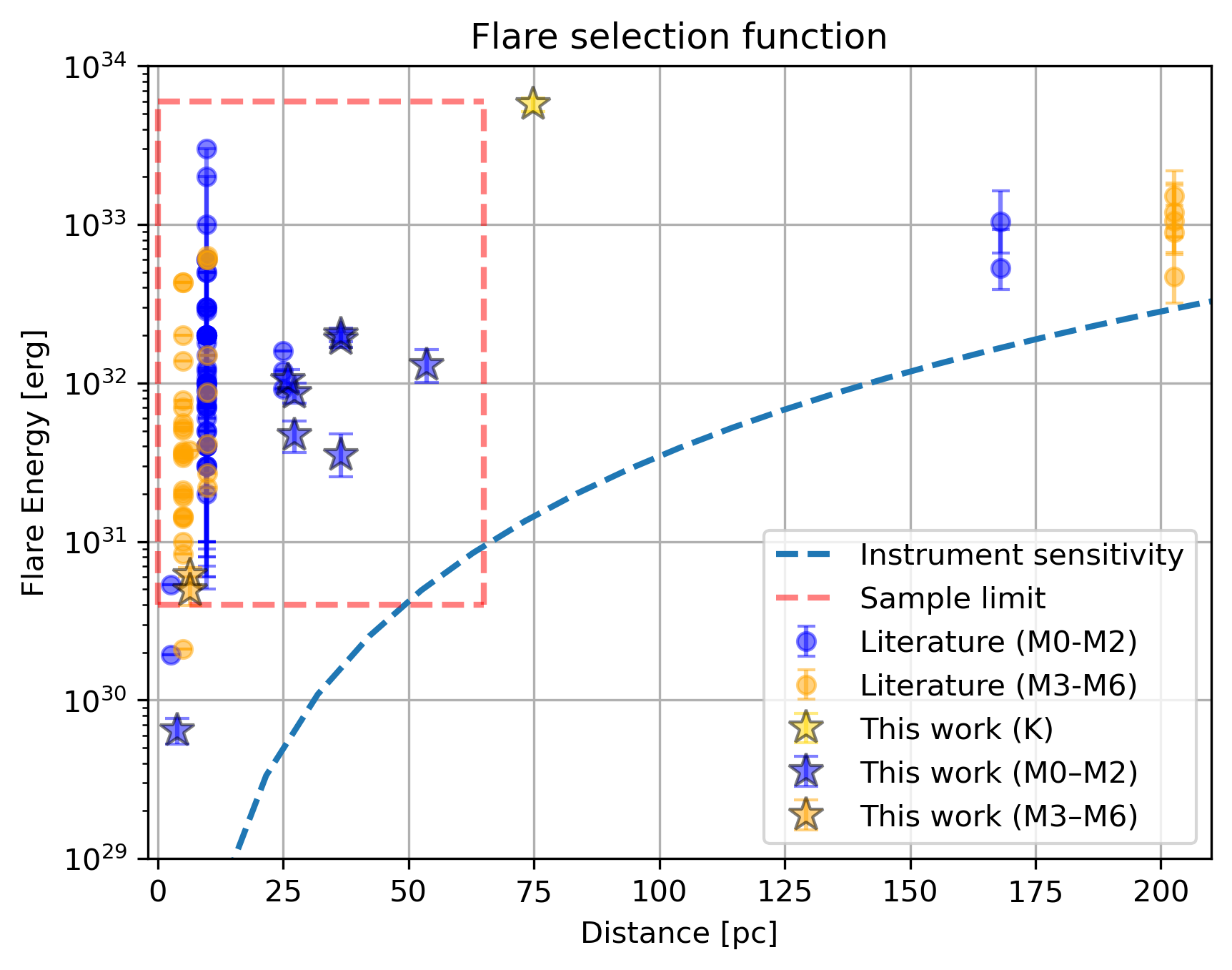}
\caption{{\bf Selecting a subsample to minimize biases.} The flaring stars analyzed in this work are shown as star markers. Those from the literature are shown as circular markers. Early M dwarfs (M0–M2) are shown in blue; mid-to-late M dwarfs (M3–M6) are shown in orange. The single K dwarf from this work is highlighted as a yellow star. The blue dashed line represents $E_{min}$, the instrument detection sensitivity as a function of distance (see equation~\ref{eq:Emin}). The dashed pink box shows the limits of the subsample on which we calculated the flare frequency distribution (FFD), selected to minimize biases. Its upper limit is set by $E_{max}$ (see \S~\ref{selection_function_flare}).}
\label{fig:flare_selection}
\end{figure*}

The minimum detectable flare energy, shown as a dotted line in Figure~\ref{fig:flare_selection} is
\begin{equation}\label{eq:Emin}
E_{\rm min}(d)=
\left(
\frac{4\pi\tau_0F_{\rm lim}}
{E_0^\beta}
\right)^{\frac{1}{1-\beta}}
d^{\frac{2}{1-\beta}}
\end{equation}
which implies $E_{\rm min}\propto d^3$ for $\beta=1/3$. For $\tau_0$ and $E_0$, we used the duration and energy measured for the Kapteyn’s Star flare (see Table~\ref{table_per_flare}), a nearby, old, and weakly active M dwarf \citep{guinan2016}.

The upper limit $E_{max}$ is constrained by the finite duration of the Chandra exposures. We assume that we are sensitive to flares with durations of approximately half the median Chandra exposure time ($\sim 20\,\rm ks$), i.e. flares with durations shorter than $\sim 10\,\rm ks$, which, according to Figure~\ref{fig:flare_energy_duration} 
corresponds to energies lower than $\sim 6\times  10^{33}\,\rm erg$.

An additional source of bias, which we do not attempt to correct for, arises from the fact that the SUPERBLINK catalog includes only the brightest and most active late-type M dwarfs as the distance grows (due to the cuts $J<10$ and $V-J>2.7$). We expect this effect to have a limited impact on our measurements because we derived the FFD separately for early-type and late-type M dwarfs, and because the late-type M dwarfs included in the FFD measurement are nearby.

The final subsample which we use to fit the FFD is shown in Figure ~\ref{fig:flare_selection}.


\section{Tabulated summary of flares properties and flaring stars properties} \label{sec:tables}

Table~\ref{table_per_flare} summarizes the properties of the flares we discovered and those reported in the literature. Table~\ref{table_per_star} summarizes the properties of the flaring M dwarfs. Both tables are available in full in the electronic version of this paper.

\begin{deluxetable*}{c c c c c c c c c c c}
\tablecaption{
\textbf{Flares properties}. The table is divided into the flaring stars discovered through the Chandra-eROSITA cross-match (upper part) and the flaring stars reported in the literature previous to this work (lower part). The SNR denotes the signal-to-noise ratio of the detection made with the methodology described in \S\,\ref{methodology}. $T_f/T_q$ is the ratio between the flare and quiescent temperatures. $E$ is the flare energy, and $F_{\rm flare}$ is the flare flux. $t_{\rm flare}$, $t_{\rm rise}$, and $t_{\rm fall}$ denote the flare duration, rise time, and decay time, respectively. $\delta$ is the equivalent duration. The full table is available in a machine-readable format with the electronic version of the paper. Additional properties, which do not appear here but are included in the electronic version of this table include the temporal FWHM for each flare, calculated both for the counts and energy light curves.} 

\label{table_per_flare}

\tablewidth{\textwidth}

\tablehead{
\colhead{\#} &
\colhead{ObsID} &
\colhead{SNR} &
\colhead{$T_f/T_q$} &
\colhead{$E$} &
\colhead{$F_{\rm flare}$} &
\colhead{$t_{\rm flare}$} &
\colhead{$t_{\rm rise}$} &
\colhead{$t_{\rm fall}$} &
\colhead{$\delta$} &
\colhead{\ldots}
\\
\colhead{} &
\colhead{} &
\colhead{} &
\colhead{} &
\colhead{[$10^{32}\,\mathrm{erg}$]} &
\colhead{[$10^{-13}\,\mathrm{erg\,s^{-1}\,cm^{-2}}$]} &
\colhead{[s]} &
\colhead{[s]} &
\colhead{[s]} &
\colhead{[s]} &
\colhead{}
}

\startdata
1&16302 & 43.712 & 5.4603 & $57.163^{+5.619}_{-5.231}$ & $37.2470^{+2.470}_{-1.497}$ & $2373.6\pm197.8$ & $395.6\pm98.9$ & $1978.0\pm98.9$ & $38006.97\pm3385.78$ & \nodata \\
2&17196 & 13.242 & 1.1559 & $1.885^{+0.222}_{-0.205}$ & $4.5912^{+0.277}_{-0.278}$ & $3218.4\pm536.4$ & $1609.2\pm268.2$ & $1609.2\pm268.2$ & $13410.00\pm1829.06$ & \nodata \\
3&17196 &  9.104 & 1.9254 & $2.024^{+0.202}_{-0.189}$ & $3.2746^{+0.203}_{-0.176}$ & $5364.0\pm536.4$ & $1072.8\pm268.2$ & $3291.2\pm268.2$ & $13410.00\pm2013.43$ & \nodata \\
4&17654 &  7.152 & 0.8746 & $0.351^{+0.124}_{-0.096}$ & $2.2751^{+0.541}_{-0.264}$ & $1537.6\pm384.4$ & $1537.6\pm192.2$ & \nodata             & $3203.33\pm1133.33$   & \nodata \\
5&20138 &  5.635 & 1.1403 & $1.301^{+0.328}_{-0.290}$ & $2.103^{+0.252}_{-0.209}$ & $3138.4\pm784.6$ & $2353.8\pm392.3$ & $784.6\pm392.3$   & $6347.97\pm1892.40$  & \nodata \\
6&21071 & 10.264 & 2.7806 & $0.0064^{+0.001}_{-0.001}$ & $5.5947^{+0.747}_{-0.716}$ & $745.0\pm149.0$ & $596.0\pm74.5$   & $149.0\pm74.5$    & $5522.93\pm1624.65$  & \nodata \\
7&24991 &  7.544 & 2.9833 & $0.867^{+0.132}_{-0.123}$ & $6.8286^{+0.548}_{-0.450}$ & $2107.2\pm351.2$ & $702.4\pm175.6$ & $1404.8\pm175.6$  & $5736.26\pm1097.50$   & \nodata \\
8&24991 &  5.956 & 1.8694 & $0.466^{+0.110}_{-0.102}$ & $4.7045^{+0.7488}_{-0.445}$ & $1756.0\pm351.2$ & $1404.8\pm175.6$ & $351.2\pm175.6$ & $3210.61\pm917.72$ & \nodata \\
9&25686 &  7.396 & 1.3927 & $1.042^{+0.178}_{-0.163}$ & $16.649^{+1.162}_{-1.001}$ & $1228.0\pm307.0$ & $921.0\pm153.5$ & $307.0\pm153.5$ & $1931.31\pm355.13$ & \nodata \\
10&27611 & 7.697 & 1.1721 & $0.060^{+0.007}_{-0.007}$ & $9.7969^{+0.760}_{-0.761}$ & $1699.2\pm212.4$ & $212.4\pm106.2$ & $1468.8\pm106.2$ & $6143.17\pm1163.19$ & \nodata \\
11&27611 & 6.670 & 1.7772 & $0.049^{+0.010}_{-0.009}$ & $8.4008^{+1.231}_{-1.099}$ & $637.2\pm212.4$  & \nodata          & $637.2\pm106.2$  & $2811.66\pm782.04$  & \nodata \\
12&624 & 11.563 &  \nodata &  \nodata & \nodata  &  $2240 \pm 448.2 $ &  $896.3 \pm 224.1 $         & $ 1344.5 \pm 224.1 $  &  \nodata & \nodata \\
\tableline
\multicolumn{11}{c}{\textbf{Total (This work): 12}}\\
\tableline
13&\nodata&\nodata&\nodata&$\sim 6$ & \nodata & 5100 & 1300 & 1700 & \nodata & \nodata \\
14&\nodata&\nodata&\nodata&$0.78$  & \nodata & 5900 & \nodata & 1730 & \nodata & \nodata \\
\ldots&\ldots&\ldots&\ldots&\ldots&\ldots&\ldots&\ldots&\ldots&\ldots&\ldots \\
\tableline
\multicolumn{11}{c}{\textbf{Total (Literature): 121}}\\
\enddata

\end{deluxetable*}

\begin{deluxetable*}{lcccccccccc} 
\tablecaption{{\bf The flaring stars sample.} The table is divided into the flaring stars discovered through the Chandra-eROSITA cross-match (upper part) and the flaring stars reported in the literature previous to this work (lower part). The Gaia source ID, equatorial coordinates (RA, Dec), and photometry for the upper part are taken from Gaia DR2 \citep{gaia2018b}, for consistency with \citet{Magaudda2022}, while values for the literature sample are taken from Gaia DR3 \citep{gaia2023}. Distance estimates ($d$) are from \citet{bailerjones2021}, based on Gaia DR3 astrometry. $n_f$ is the number of flares detected for each star.
We note that LP 944–20 was previously reported in the literature as an X-ray flaring source; it was not discovered by us, but we performed the calculations of the parameters shown in the table.
In the literature sample, stellar names follow those reported by the authors in the Reference column (see machine-readable version of the table). Stars with two Gaia source IDs (e.g., AT~Mic) are resolved binaries. Empty entries indicate quantities not provided in the cited reference.}
\label{table_per_star}

\tablehead{
\colhead{Simbad Name} &
\colhead{Gaia Source ID} &
\colhead{RA} &
\colhead{Dec} &
\colhead{$G_{\rm mag}$} &
\colhead{$B_P-R_P$} &
\colhead{d} &
\colhead{$t_e$} &
\colhead{$t_f$} &
\colhead{$n_f$}
\\
\colhead{} & \colhead{} & \colhead{[deg]} & \colhead{[deg]} &
\colhead{[mag]} & \colhead{[mag]} & \colhead{[pc]} &
\colhead{[ks]} & \colhead{[ks]} & \colhead{}
}

\startdata
                   $PM J13569-3236$& 6170841197931246592 & 209.23094962301 & -32.60968999131 & 11.2511 $\pm$0.001  & 1.5529 & 74.80 $\pm$ 2.44 & 39.7316 & 2.3736 & 1 \\
                    $PM J13155-1627$& 3511397894924621952 & 198.88209170913 & -16.45658596211 & 11.5832 $\pm$ 0.0014 & 2.0831 & 36.56 $\pm$ 0.05 & 342.6424 & 10.1200 & 3 \\
                    $PM J10431+2218$& 3989472560067645056 & 160.78229336548 & 22.30176832517 & 12.1971 $\pm$ 0.0011 & 2.0491 & 53.64 $\pm$ 0.05 & 37.1054 & 3.1384 & 1 \\
                   HD 33793 (Kapteyn's star) & 4810594479417465600 & 77.95866129564 & -45.04301982784 & 8.062 $\pm$ 0.0006 & 2.0336 & 3.93 $\pm$ 0.01 & 62.5144 & 0.7450 & 1 \\
                    $V^{*} KX Com$& 3956837478702437632 & 194.21998779982 & 23.49740493257 & 12.6686 $\pm$ 0.0014 & 2.5929 & 27.31 $\pm$ 0.02 & 64.1893 & 3.8632 & 2 \\
                    $	
VSS VI-20$& 3219132482897057152 & 86.82454194 & -0.01385605654 & 10.2111 $\pm$ 0.0012 & 1.9235 & 25.91 $\pm$ 0.01 & 10.0901 & 1.2280 & 1 \\
                    $V^{*} GL Vir$& 3907283967108054528 & 184.74192716159 & 11.12692402507 & 11.9366 $\pm$ 0.0009 & 3.5475 & 6.46 $\pm$ 0.00 & 20.0495 & 2.3364 & 2 \\
                    \hline
\multicolumn{7}{c}{\textbf{Total (This work): 7}}&576.323&23.804&11 \\
\hline
LP 944--20          & 4860376345833699840 & 54.89857021  & $-35.42759364$ & 15.440 $\pm$ 0.0029 & 5.057 & 6.423 $\pm$ 0.002  & 43.8170 & 2.2409 & 1 \\
UV Ceti            & 5140693571158946048 & 24.77167421  & $-17.94768286$ & 10.818 $\pm$ 0.0034 & 3.829 & 2.675 $\pm$ 0.003   & 30.2400 &  & 1 \\
kic\_8093473       & 2126861693247227136 & 290.36867151 & 43.92934159   & 14.693 $\pm$ 0.0034 & 2.425 & 202.67 $\pm$ 3.462  &  &  &6 \\
kic\_8454353       & 2079043107926784384 & 299.09977222 & 44.49169642   & 14.915 $\pm$ 0.0029 & 2.226 & 167.96  $\pm$ 0.681 &  &  &2 \\
AD Leo             & 625453654702751872  & 154.89881370 & 19.86980991   & 8.204 $\pm$ 0.0028  & 2.565 & 4.964 $\pm$ 0.001   & 191.8958 & 35.7300 & 16 \\
EV Lac             & 1934263333784036736 & 341.70282545 & 44.33195332   & 9.005 $\pm$ 0.0028 & 2.730 & 5.050 $\pm$ 0.001   & 159.9186 & 73.1920 & 18 \\
AU Mic             & 6794047652729201024 & 311.29118265 & $-31.34250008$ & 7.843 $\pm$ 0.0028 & 2.111 & 9.710 $\pm$ 0.002  & 544.0631 & 186.3498 & 46 \\
YZ Cmi             & 3136952686035250688 & 116.16583593 & 3.55048445    & 9.692 $\pm$ 0.0029 & 2.997 & 5.987 $\pm$ 0.001  & 46.1179 & 3.5300 & 5 \\
kic\_9048551       & 2079184429529920768 & 299.50491648 & 45.30163256   & 13.328 $\pm$ 0.0029 & 1.749 & 125.42 $\pm$ 0.091 &  &  &3 \\
HD 95735           & 762815470562110464  & 165.83095968 & 35.94865303   & 6.551 $\pm$ 0.0028 & 2.216 & 2.546 $\pm$ 0.000  & 40.7007 & 7.3000 & 2 \\
DG Cvn             & 1450067137649786752 & 202.94302916 & 29.27619084   & 10.628 $\pm$ 0.0028 & 2.894 &     & 31.923400 & 4.9400 & 4 \\
HK Aqr             & 2410011260521088512 & 347.08195599 & $-15.41001897$ & 10.186 $\pm$ 0.0033 & 1.942 & 24.92 $\pm$ 0.009  & 74.2040 & 9.0600 & 3 \\
EQ Peg A           & 2824770686019003904 & 352.97012375 & 19.93699283   & 9.044 $\pm$ 0.0028 & 2.661 & 6.261 $\pm$ 0.001  & 39.2888 & 8.3800 & 2 \\
EQ Peg B                  & 2824770686019004032 & 352.97167530 & 19.93731375   & 10.829 $\pm$ 0.0028 & 3.174 & 6.251 $\pm$ 0.002  &  &  &  \\
AT Mic A         & 6792436799475128960 & 310.46433403 & $-32.43703678$ & 9.576 $\pm$ 0.0029 & 3.031 & 9.915 $\pm$ 0.007  &  &  &  \\
AT Mic B           & 6792436799477051904 & 310.46473122 & $-32.43751492$ & 9.605 $\pm$ 0.0028 & 3.126 & 9.802 $\pm$ 0.007  & 104.7921 & 21.6400 & 11 \\
V645 (Proxima) Cen  & 5853498713190525696 & 217.39232147 & $-62.67607511$ & 8.985 $\pm$ 0.0028 & 3.805 & 1.301 $\pm$ 0.000  & 94.8800 & 2.4000  & 2 \\
\hline
\multicolumn{7}{c}{\textbf{Total (Literature): {17 (of which $3$ binaries)}}}& 1401.848&354.763&122\\
\enddata
\end{deluxetable*}

\begin{table*}[ht]
\centering
\caption{Flare Energies and Fluxes}
\label{tab:energy_table}

\resizebox{\textwidth}{!}{
\begin{tabular}{ccccccccc}
\toprule
\hline\hline
\multicolumn{1}{c}{} & 
\multicolumn{2}{c}{Energy} & 
\multicolumn{3}{c}{Quiescent Flux} & 
\multicolumn{3}{c}{Flare Flux} \\
\cline{2-3}\cline{4-6}\cline{7-9}
Flare Index & Median & 16--84\% & Best Fit & Median & 16--84\% & Best Fit & Median & 16--84\% \\
 & 
\multicolumn{2}{c}{(erg)} & 
\multicolumn{3}{c}{(erg\,cm$^{-2}$\,s$^{-1}$)} & 
\multicolumn{3}{c}{(erg\,cm$^{-2}$\,s$^{-1}$)} \\
\midrule
\hline
1 & $5.72 \times 10^{33}$ & $(5.19\text{--}6.28) \times 10^{33}$ & $1.684 \times 10^{-13}$ & $1.683 \times 10^{-13}$ & $(1.59\text{--}1.80) \times 10^{-13}$ & $3.725 \times 10^{-12}$ & $3.776 \times 10^{-12}$ & $(3.58\text{--}3.97) \times 10^{-12}$ \\
2 & $1.89 \times 10^{32}$ & $(1.68\text{--}2.11) \times 10^{32}$ & $9.260 \times 10^{-14}$ & $9.235 \times 10^{-14}$ & $(8.80\text{--}9.68) \times 10^{-14}$ & $4.591 \times 10^{-13}$ & $4.596 \times 10^{-13}$ & $(4.31\text{--}4.87) \times 10^{-13}$ \\
3 & $2.02 \times 10^{32}$ & $(1.83\text{--}2.23) \times 10^{32}$ & $9.260 \times 10^{-14}$ & $9.235 \times 10^{-14}$ & $(8.80\text{--}9.68) \times 10^{-14}$ & $3.275 \times 10^{-13}$ & $3.284 \times 10^{-13}$ & $(3.10\text{--}3.48) \times 10^{-13}$ \\
4 & $3.51 \times 10^{31}$ & $(2.56\text{--}4.76) \times 10^{31}$ & $9.239 \times 10^{-14}$ & $9.186 \times 10^{-14}$ & $(8.49\text{--}9.96) \times 10^{-14}$ & $2.275 \times 10^{-13}$ & $2.352 \times 10^{-13}$ & $(2.01\text{--}2.82) \times 10^{-13}$ \\
5 & $1.30 \times 10^{32}$ & $(1.01\text{--}1.63) \times 10^{32}$ & $8.993 \times 10^{-14}$ & $8.995 \times 10^{-14}$ & $(8.50\text{--}9.52) \times 10^{-14}$ & $2.103 \times 10^{-13}$ & $2.116 \times 10^{-13}$ & $(1.89\text{--}2.36) \times 10^{-13}$ \\
6 & $6.41 \times 10^{29}$ & $(5.29\text{--}7.69) \times 10^{29}$ & $9.413 \times 10^{-14}$ & $9.066 \times 10^{-14}$ & $(7.78\text{--}10.6) \times 10^{-14}$ & $5.595 \times 10^{-13}$ & $5.578 \times 10^{-13}$ & $(4.88\text{--}6.34) \times 10^{-13}$ \\
7 & $8.67 \times 10^{31}$ & $(7.43\text{--}9.99) \times 10^{31}$ & $2.235 \times 10^{-13}$ & $2.225 \times 10^{-13}$ & $(2.00\text{--}2.48) \times 10^{-13}$ & $6.829 \times 10^{-13}$ & $6.855 \times 10^{-13}$ & $(6.38\text{--}7.38) \times 10^{-13}$ \\
8 & $4.66 \times 10^{31}$ & $(3.63\text{--}5.76) \times 10^{31}$ & $2.235 \times 10^{-13}$ & $2.225 \times 10^{-13}$ & $(2.00\text{--}2.48) \times 10^{-13}$ & $4.705 \times 10^{-13}$ & $4.722 \times 10^{-13}$ & $(4.26\text{--}5.19) \times 10^{-13}$ \\
9 & $1.04 \times 10^{32}$ & $(8.80\text{--}12.2) \times 10^{31}$ & $6.117 \times 10^{-13}$ & $6.119 \times 10^{-13}$ & $(5.83\text{--}6.42) \times 10^{-13}$ & $1.665 \times 10^{-12}$ & $1.675 \times 10^{-12}$ & $(1.56\text{--}1.78) \times 10^{-12}$ \\
10 & $6.09 \times 10^{30}$ & $(5.37\text{--}6.88) \times 10^{30}$ & $2.606 \times 10^{-13}$ & $2.588 \times 10^{-13}$ & $(2.43\text{--}2.77) \times 10^{-13}$ & $9.797 \times 10^{-13}$ & $9.768 \times 10^{-13}$ & $(9.04\text{--}10.6) \times 10^{-13}$ \\
11 & $4.93 \times 10^{30}$ & $(3.95\text{--}6.00) \times 10^{30}$ & $2.606 \times 10^{-13}$ & $2.588 \times 10^{-13}$ & $(2.43\text{--}2.77) \times 10^{-13}$ & $8.401 \times 10^{-13}$ & $8.410 \times 10^{-13}$ & $(7.30\text{--}9.63) \times 10^{-13}$ \\
\hline
\bottomrule
\end{tabular}
}
\end{table*}

\clearpage
\bibliography{bibliograph}
\bibliographystyle{aasjournalv7}

\end{document}